\documentclass[pdflatex,sn-mathphys-num]{sn-jnl}

\usepackage{graphicx}%
\usepackage{multirow}%
\usepackage{amsmath,amssymb,amsfonts}%
\usepackage{amsthm}%
\usepackage{mathrsfs}%
\usepackage[title]{appendix}%
\usepackage{xcolor}%
\usepackage{textcomp}%
\usepackage{manyfoot}%
\usepackage{booktabs}%
\usepackage{algorithm}%
\usepackage{algorithmicx}%
\usepackage{algpseudocode}%
\usepackage{listings}%

\usepackage{booktabs}
\usepackage{makecell}
\usepackage{tabularx}
\usepackage{enumitem}

\newcolumntype{C}{>{\centering\arraybackslash}m{5cm}}

\theoremstyle{thmstyleone}%
\newtheorem{theorem}{Theorem}
\newtheorem{proposition}[theorem]{Proposition}%
\newtheorem{claim}{Claim}
\newtheorem{lemma}{Lemma}

\theoremstyle{thmstyletwo}%
\newtheorem{example}{Example}%

\theoremstyle{thmstylethree}%
\newtheorem{definition}{Definition}%

\newcommand{\mset}[1]{\{\!\!\{ #1 \}\!\!\}}

\DeclareMathOperator{\sign}{sign}

\begin{document}

\title[Aggregating Rankings of Coalitions to Rank Individuals]{Aggregating Rankings of Coalitions to Rank Individuals}


\author*[1]{\fnm{Takahiro} \sur{Suzuki}}\email{suzuki-tkenmgt@g.ecc.u-tokyo.ac.jp}

\affil*[1]{\orgdiv{Department of Civil Engineering}, \orgname{The University of Tokyo}, \orgaddress{\street{7-3-1}, \city{Hongo, Bunkyo-ku}, \postcode{113-8656}, \state{Tokyo}, \country{Japan}}}


\abstract{
We develop an ordinal framework for aggregating a profile of rankings of coalitional performances to determine the ranking of individuals. From university ratings to the assessment of team-sports players, we often evaluate individuals based on their coalitional performances. Although such evaluation has been addressed by cooperative game theory, it is only recently that its ordinal counterpart, \textit{social ranking} theory \cite{morettiAxiomaticApproachSocial}, has emerged. As most of the literature considers a single coalitional ranking (which corresponds to a real-valued characteristic function), we extend the model to allow a profile of coalitional rankings (which corresponds to a multidimensional characteristic function). We tackle the aggregation problem by introducing three families of inter-profile axioms, each representing how the rankings should interact or contribute to the output. After revealing fundamental impossibilities concerning the reconciliation of different families, we characterize a new solution, Pareto-extension Lex-cel (PEL), via some weaker versions of the inter-profile axioms. PEL extends Lex-cel, which is a central solution in social ranking theory, by analogy with Sen's Pareto extension rule. Our results develop the connection between classical Arrovian aggregation, cooperative game theory, and social ranking theory.}




\maketitle

\section{Introduction}\label{section:introduction}
We examine the general problem of aggregating multiple rankings of coalitions (e.g., based on multiple criteria for coalitional performance or supplied by referees or data sources) to determine the ranking of individuals. University rankings such as THE, QS, and ARWU, for instance, evaluate the quality of universities using various indicators such as ``citations per faculty'', ``international faculty'', and ``Alumni, Nobel/Fields''. Notably, some of these indicators measure individual characteristics (e.g., ``international faculty''), but others measure cooperative achievements across several universities (e.g., in relation to citations, the authors of a paper are often affiliated with multiple universities). 

As an ordinal counterpart to cooperative game theory, the theory of \textit{social ranking} \cite{morettiAxiomaticApproachSocial,morettiAxiomaticAlgorithmicPerspectives2017} has recently emerged to evaluate individuals. Social ranking theory typically analyzes the mapping from a ``single'' ranking of coalitions (\textit{power relation}) to the ranking of individuals (\textit{social ranking}). A typical case is the evaluation of researchers based on their collaborative achievements (Example \ref{example:researchers}):

\begin{example}[Social ranking]
\label{example:researchers}
Consider a situation in which four researchers ($1,2,3$, and $4$) are to be ranked. According to their achievements in economics journals, $\{1,3\}$ (i.e., joint work by $1$ and $3$) is the most productive, $\{2,3\}$ and $\{1,2\}$ are tied as the second most productive, and $\{3\}$ and $\{1,4\}$ are the least productive. Hence, the economics ranking $\succsim^E$ is described as follows:\footnote{As usual, $\succ^E$ represents the asymmetric part of $\succsim^E$ ($A\succ^E B$ means $A$ is strictly better than $B$); $\sim^E$ represents the symmetric part of $\succsim^E$ ($A\sim^E B$ means $A$ is indifferent to $B$).}
\[\{1,3\}\succ^E\{2,3\}\sim^E\{1,2\}\succ^E\{3\}\sim^E\{1,4\}.\]
Then, what would be an appropriate ranking of $1,2,3$, and $4$ in this case?
\end{example}

The main purpose of this paper is to extend the existing theory of social ranking and develop an ordinal framework to aggregate a profile of multiple power relations in such evaluations. In many practical contexts such as the evaluation of researchers, the abundance of performance criteria is fundamental. In fact, researchers are often evaluated from various perspectives such as publications, teaching skills, and acquisition of competitive funding. 

Formally, we study the mapping, which we call a \textit{social ranking solution} (SRS), from a profile of power relations to a ranking of individuals. Focusing on the relationship between the (profile of) power relations---for instance, whether they reinforce each other or how multiple sources can be integrated---we introduce three families of ``inter-profile'' axioms. Let us illustrate them using an example:

\begin{example}
\label{example:researchers2} 
In addition to the power relation $\succsim^E$ in Example \ref{example:researchers}, which reflects the ranking with respect to economics, assume that power relations $\succsim^M$ (ranking based on publications in mathematics) and $\succsim^T$ (ranking with respect to teaching skills) are also given. The profile $\pi$ of these three power relations is given as follows: 

\[\pi:\begin{cases}
\{1,3\}\succ^E\{2,3\}\sim^E\{1,2\}\succ^E\{3\}\sim^E\{1,4\}\\
\{3\}\succ^M\{1\}\succ^M\{2\}\succ^M\{3,4\}\\
\{2,3\}\succ^T\{1,3\}\succ^T\{1\}\sim^T\{2\}.
\end{cases}\]
Then, again, what is a reasonable ranking of individuals $1$, $2$, $3$, and $4$?
\end{example}

We consider three families of inter-profile axioms: \textit{cut} (CUT), consistency (CON), and union consistency (UCON). The intuition behind them is as follows. 

\begin{enumerate}
\item CUT focuses on the integration of two power relations via some `couplers'. For instance, in Example \ref{example:researchers2}, the economics ranking $\succsim^E$ provides a ground for considering coalitions to be at least as good as $\{3\}$, while the mathematics ranking $\succsim^M$ describes coalitions at least as bad as $\{3\}$. Hence, if we simply integrate\footnote{Notice that CUT regards each ranking as a ground for considering some coalitions to be at least as good (bad) as the coupler, and focuses on the number of such grounds to support the claim. This idea is argued more generally in the opinion aggregation model \cite{barberaOpinionAggregationBorda2023}.} these two rankings with $\{3\}$ as the coupler, we can represent $\pi$ as follows ($\succsim^R$ is the integrated ``research'' ranking):
\[\pi':\begin{cases}
\{1,3\}\succ^R\{2,3\}\sim^R\{1,2\}\succ^R\{3\}\sim^R\{1,4\}\succ^R\{1\}\succ^R\{2\}\succ^R\{3,4\},\\
\{2,3\}\succ^T\{1,3\}\succ^T\{1\}\sim^T\{2\}
\end{cases}\]
\item CON focuses on the decomposition of power relations; if we partition a profile $\pi$ into two profiles $\pi_1$ and $\pi_2$ by partitioning each power relation, CON demands that the social rankings obtained at $\pi_1$ and $\pi_2$ should be coherent with the social ranking obtained at $\pi$. For instance, in Example \ref{example:researchers2}, suppose that only $3$ is a senior researcher. Then, it would be reasonable to partition $\pi$ into $\pi_1$ (coalitions made up of only young researchers) and $\pi_2$ (coalitions including the senior researcher, $3$) as follows: 
\[\pi_1:\begin{cases}
\{1,2\}\succ^{E'}\{1,4\}\\
\{1\}\succ^{M'}\{2\}\\
\{1\}\sim^{T'}\{2\}
\end{cases}\text{ and }
\pi_2:\begin{cases}
\{1,3\}\succ^{E''}\{2,3\}\succ^{E''}\{3\}\\
\{3\}\succ^{M''}\{3,4\}\\
\{2,3\}\succ^{T''}\{1,3\}.
\end{cases}\]
Notice that each $\succsim_A$ (where $A\in \{E,M,T\}$) is partitioned into $\succsim^{A'}$ and $\succsim^{A''}$. CON requires that, for each $x,y\in \{x,y\}$, if the social ranking of $x$ is no worse than that of $y$ at both $\pi_1$ and $\pi_2$, then this should also hold at $\pi$ (and if the social ranking of $x$ is strictly better at either $\pi_1$ or $\pi_2$, then this should also hold at $\pi$). Our CON is a direct extension of the axiom of the same name in the theory of social ranking \cite{suzukiConsistentSocialRanking2024,ruelleCeterisParibusBorda2026,suzukiConsistenciesSocialRanking2026}.
\item UCON focuses on the decomposition of the profile into two parts. For instance, in the case of Example \ref{example:researchers2}, if the social ranking of $x$ is no worse than [resp. strictly better than] that of $y$ at the following profile (i.e., a sub-profile with respect to research)

\[\pi_3:\begin{cases}
\{1,3\}\succ^E\{2,3\}\sim^E\{1,2\}\succ^E\{3\}\sim^E\{1,4\}\\
\{3\}\succ^M\{1\}\succ^M\{2\}\succ^M\{3,4\}\\
\end{cases}\]

and the social ranking of $x$ is no worse than that of $y$ at the following profile

\[\pi_4:\begin{cases}
\{2,3\}\succ^T\{1,3\}\succ^T\{1\}\sim^T\{2\},
\end{cases}\]
then the social ranking of $x$ should be no worse [resp. strictly better] than that of $y$ at $\pi$ as well. UCON is a direct extension of the consistency axiom in voting theory \cite{youngAxiomatizationBordasRule1974,nitzanFurtherCharacterizationBorda1981,debordAxiomaticCharacterizationBordas1992,alos-ferrerSimpleCharacterizationApproval2006}.\footnote{As we consider a relation-valued function, the closest study among those cited is \cite{nitzanFurtherCharacterizationBorda1981}.}

\end{enumerate}

Our contribution is mainly twofold. First, we will prove a series of fundamental impossibilities concerning the different families of inter-profile axioms (Section \ref{section:multiple inter-profile axioms}). We will show that any pair of the three families of inter-profile axioms, when reconciled, leads to an impossibility result under some natural axioms such as neutrality. Second, we focus on how to extend the \textit{lexicographic excellence solution} (Lex-cel) in our setting. Among the many SRSs studied so far, Lex-cel is a solution that compares individuals lexicographically by counting the number of their coalitions ranked in the highest class, second class, and so on. We first show that, no matter which inter-profile axiom (CUT, CON, or UCON) we apply, we cannot extend Lex-cel to our setting while preserving its well-known basic properties (Section \ref{section: further impossibilities}). As an escape route from these impossibilities, we introduce a new solution, called Pareto-extension Lex-cel (PEL). We characterize PEL using some weaker versions of the inter-profile axioms (Section \ref{section:characterization of PEL}). Notice that PEL is named after Sen's Pareto extension rule \cite{senQuasiTransitivityRationalChoice1969}. Sen's Pareto extension rule is a classical escape route from Arrow's impossibility by weakening social rationality. In fact, our aggregation problem has much in common with the classical Arrovian aggregation problem, and that partly explains why Sen's approach performs well in our model. A detailed discussion of these two models is provided in Section \ref{subsection:discussion}.

\paragraph{Related literature}
The present paper directly contributes to the theory of social ranking \cite{morettiAxiomaticApproachSocial,morettiAxiomaticAlgorithmicPerspectives2017} (see, e.g., \cite{lucchettiConnectingPreferencesSet2026} for a recent review). In the last decade, many SRSs have been proposed and axiomatically studied: plurality \cite{suzukiMillsCanonsMeet2025}, the CP Borda rule \cite{ruelleCeterisParibusBorda2026}, the Borda rule \cite{suzukiConsistencyUnanimityBorda2026}, CP majority \cite{haretCeterisParibusMajority2018}, intersection initial segment \cite{suzukiRankingAlternativesOpinions2026}, the lexicographic excellence solution (Lex-cel) and its variant \cite{bernardiRankingObjectsPreference2019c,algabaLexicographicSolutionsCoalitional2021}, and the ordinal Banzhaf index \cite{khaniOrdinalBanzhafIndex2019}. Furthermore, computer-aided simulations concerning manipulability \cite{alloucheSocialRankingManipulability2020b} or explainable AI \cite{gourvesFeatureSelectionLexicographic2026} have also been conducted. With only a few exceptions\footnote{\cite{suzukiSocialRankingProblem2021} studies the extension of the results by \cite{morettiAxiomaticAlgorithmicPerspectives2017} when there are multiple indicators; \cite{suzukiRankingAlternativesOpinions2026,suzukiAggregatingOpinionsSets2025} study a highly abstract model in which only the number of opinions matters.}, most of the existing research has dealt with the case of a single power relation as the input. Our main contribution is to introduce a generalized model in which a profile (list) of power relations is taken into consideration. This allows us to model multi-criteria situations. From this perspective, another interpretation of our research is to fill in the gap between the theory of social ranking and multi-criteria decision making (MCDM) \cite{dealmeidaMulticriteriaMultiobjectiveModels2015}. In the MCDM context, a certain lexicographic method (e.g., pp. 74--75 in \cite{hwangMultipleAttributeDecision1981}) is a classical method with some recent application \cite{nurmiChoiceVotingRules2015}. 

Our study is also related to cooperative game theory. In this century, various refinements have been made to the basic model of a transferable utility game (see, e.g., \cite{branzeiCooperativeIntervalGames2010} for a survey). Among them, our study is closely related to research focusing on the refinement of characteristic functions (e.g., partially ordered \cite{puertoPartiallyOrderedCooperative2008}; vector-valued \cite{fernandezCoreSolutionsVectorValued2002}; multiple-scenario \cite{hinojosaCoreLeastCore2005}). In relation to this line of research, our ordinal approach enlarges the scope of the research by enabling us to deal with qualitative criteria that are hard to measure on a cardinal scale. 

\paragraph{Roadmap of the paper}
Section \ref{section:model} provides the basic model, including the definitions of inter-profile and complementary axioms. Our main results are presented in Sections \ref{section:multiple inter-profile axioms} and \ref{section:extending Lex-cel}. The former describes the impossibilities concerning multiple families of inter-profile axioms, and the latter examines extensions of Lex-cel. Concluding remarks are in Section \ref{section:conclusion}. Most of the proofs appear in the Appendix; the others are given in the main text. 

\section{Model}
\label{section:model}

\subsection{Basic model}
\label{subsection:basics}
\paragraph{Preliminaries}
For any set $A$, a \textit{binary relation} on $A$ is a subset of $A\times A$. For a binary relation $R\subseteq A\times A$ and $a,b\in A$, we write $aRb$ instead of $(a,b)\in R$. A binary relation $R$ on $A$ is called \textit{complete} if $aRa$ holds for all $a\in A$; \textit{transitive} if for any $a,b,c\in A$, we have $[(aRb\land bRc)\Rightarrow aRc]$. A binary relation is called a \textit{weak order} if it is complete and transitive. We denote by $\mathcal{B}(A)$ the set of all complete binary relations on $A$, and by $\mathcal{R}(A)$ the set of all weak orders on $A$.

We denote by $\mathbb{N}=\{0,1,2,\dots\}$ the set of all nonnegative integers, and by $\mathbb{N}_{+}=\{1,2,\dots\}$ the set of all positive integers. For each $n\in \mathbb{N}$, let $[n]:=\{a\in \mathbb{N}\mid a\leq n\}$ denote the set of all positive integers equal to or less than $n$.

\paragraph{Model}
Let $X$ be the set of all individuals. We assume that $X$ is countably infinite. Let $\mathcal{X}$ be the set of all finite subsets of $X$, i.e., $\mathcal{X}:=\{S\subseteq X\mid \lvert S\rvert<\infty\}$. Each element of $\mathcal{X}$ is called a \textit{coalition}. Let $\mathcal{R}:=\sum_{\substack{\mathcal{D}\subseteq \mathcal{X}:\\ \lvert \mathcal{D}\rvert<\infty}}\mathcal{R}(\mathcal{D})$ denote the set of all weak orders on some finite subset $\mathcal{D}$ of $\mathcal{X}$. Each element of $\mathcal{R}$ is called a \textit{power relation}. For each $\succsim\in \mathcal{R}(\mathcal{D})$, $\mathcal{D}$ is interpreted as the set of all feasible coalitions, and $\succsim$ represents a performance ranking of the feasible coalitions. Let $\Pi:=\{\pi:\mathcal{R}\rightarrow \mathbb{N}\cup\{0\}\mid \sum_{\succsim\in \mathcal{R}}\pi(\succsim)<\infty\}$. Each element $\pi\in \Pi$ is called a \textit{profile} (of power relations). For a profile $\pi$ and $\succsim\in \mathcal{R}$, we interpret $\pi(\succsim)$ as how many indicators yield the ranking $\succsim$. Therefore, an alternative representation of a profile $\pi$ is a multiset in which each $\succsim$ appears $\pi(\succsim)$ times (i.e., $\pi(\succsim)$ is the multiplicity of $\succsim$). For instance, suppose that three indicators yield the ranking $\succsim_1$ (e.g., according to economics, mathematics, and awards, the researchers' coalitions are ranked as $\succsim_1$), and two indicators yield the ranking $\succsim_2$ (e.g., according to competitive funding and class reputations, researchers' coalitions are ranked as $\succsim_2$). Then, this situation is expressed either as (i) [functional style] $\pi:\mathcal{R}\rightarrow \mathbb{N}\cup\{0\}$ such that $\pi(\succsim_1)=3$, $\pi(\succsim_2)=2$, and $\pi(\succsim)=0$ for any other $\succsim$, or (ii) [multiset style] $\pi=\mset{\succsim_1,\succsim_1,\succsim_1,\succsim_2,\succsim_2}$ (to distinguish sets from multisets, we use $\{\}$ for ordinary sets and $\mset{}$ for multisets). An arbitrary profile is often denoted as $\pi=\mset{\succsim_1,\succsim_2,\dots,\succsim_n}$, where $\succsim_i\in \mathcal{R}$ and some of $\succsim_1,\succsim_2,\dots,\succsim_n$ can be identical. For $\pi\in\Pi$, the \textit{cardinality} of $\pi$ is defined as $\lvert\pi\rvert:=\sum_{\succsim\in\mathcal{R}}\pi(\succsim)$. 

\begin{definition}
A \textit{social ranking solution} (SRS) is a function $R:\Pi\rightarrow \mathcal{B}(X)$.
\end{definition}

An SRS $R$ maps a profile $\pi$ of power relations to a complete binary relation $R(\succsim)$ on $X$. Our domain $\Pi$ of the SRS is broader than the standard domain, which is $\mathcal{R}$ in many studies, in that we allow the case of multiple power relations (indicators such as criteria or reviewers). Such multiplicity should be essential in many practical evaluations of universities, researchers, or team-sports athletes, as their roles/objectives are often diverse. It is noteworthy that we treat each indicator equally: we do not consider the identity of each indicator; each is simply counted as a power relation. 

For convenience, we will introduce some further notational conventions. Let $\pi,\pi_1,\pi_2,\dots,\pi_n\in \Pi$ be profiles, $\succsim\in \mathcal{R}(\mathcal{D})$ be a power relation, and $R$ be an SRS.  
\begin{enumerate}[label=(\roman*)]
\item We denote $R(\pi)$ as $R_{\pi}$. If $\pi$ is a singleton (i.e., $\pi=\mset{\succsim}$ for some $\succsim\in \mathcal{R}$), we often denote $R_{\mset{\succsim}}$ as $R_{\succsim}$.
\item As usual, the asymmetric part (resp. symmetric part) of $R_{\succsim}$ is denoted as $P_{\succsim}$ (resp. $I_{\succsim}$). 
\item The \textit{sum} $\pi_1+\pi_2$ of $\pi_1$ and $\pi_2$ is defined as $(\pi_1+\pi_2)(\succsim):=\pi_1(\succsim)+\pi_2(\succsim)$ for all $\succsim\in \mathcal{R}$. The summation symbol $\sum$ is also used for the sum of more than two profiles: $\Sigma_{k\in[n]}\pi_{k}:=((\pi_1+\pi_2)+\pi_3)\dots+\pi_n$.\footnote{Since $+$ is trivially associative, the order of summation does not matter.}
\item For $\mathcal{D}'\subseteq \mathcal{D}$, the restriction of $\succsim$ to $\mathcal{D}'$ is denoted as $\succsim\mid_{\mathcal{D}'}$, i.e., $\succsim\mid_{\mathcal{D}'}:=\{(S,T)\in \succsim\mid S,T\in \mathcal{D}'\}$.
\end{enumerate}

For $\succsim\in \mathcal{R}(\mathcal{D})$, we write $\succsim:\Sigma_1\succ\Sigma_{2}\succ\cdots\succ\Sigma_{l}$ (this is often called a quotient order of $\succsim$) if the following conditions 1--3 hold: 
\begin{enumerate}
\item the set of all greatest elements of $\mathcal{D}$ with respect to $\succsim$ is $\Sigma_{1}$ (i.e., $\Sigma_{1}=\{S\in \mathcal{D}\mid S\succsim T\text{ for all }T\in \mathcal{D}\}$),
\item for each $k=2,3,\dots,n$, $\Sigma_{k}$ is the set of all greatest elements of $\bigcup(\mathcal{D}\setminus (\Sigma_{1}\cup\dots\cup\Sigma_{k-1}))$ with respect to $\succsim$, and 
\item $\Sigma_{1}\cup\Sigma_{2}\cup\dots\cup\Sigma_{l}=\mathcal{D}$.
\end{enumerate}
In words, each $\Sigma_{k}$ ($k\in[l]$) is an equivalence class with respect to $\succsim$, and $\Sigma_{k}$ is the $k$-th best equivalence class. If all elements in $\mathcal{D}$ are indifferent under $\succsim$, then it follows that $\succsim:\mathcal{D}$. In such a case, $\succsim$ is called \textit{single-class}. According to this convention, we can express the power relations discussed in Example \ref{example:researchers2} and the subsequent argument as in Example \ref{example:researchers3}. Notice that irrespective of the symbols of the power relations (e.g., $\succsim^E,\succsim^M,\succsim^T,\succsim^R$), we place $\succ$ between the equivalence classes.

\begin{example}
\label{example:researchers3}
\begin{align*}
\succsim^E&:\{\{1,3\}\}\succ\{\{2,3\},\{1,2\}\}\succ\{\{3\},\{1,4\}\},\\
\succsim^M&:\{\{3\}\}\succ\{\{1\}\}\succ\{\{2\}\}\succ\{\{3,4\}\},\\
\succsim^T&:\{\{2,3\}\}\succ\{\{1,3\}\}\succ\{\{1\},\{2\}\},\\
\succsim^R&:\{\{1,3\}\}\succ\{\{2,3\},\{1,2\}\}\succ\{\{3\},\{1,4\}\}\succ\{\{1\}\}\succ\{\{2\}\}\succ\{\{3,4\}\}
\end{align*}
\end{example}

Furthermore, a profile $\pi:=\mset{\succsim_1,\succsim_2,\dots,\succsim_n}$ with $\succsim_i:\Sigma^i_1\succ\Sigma^i_2\succ\dots\succ\Sigma^i_n$ ($1\leq i\leq l_i$) is often denoted as
\[\pi:\begin{cases}
\succsim_1:\Sigma^1_1\succ\Sigma^1_2\succ\dots\succ\Sigma^1_{l_1}\\
\succsim_2:\Sigma^2_1\succ\Sigma^2_2\succ\dots\succ\Sigma^2_{l_2},\\
\vdots\\
\succsim_n:\Sigma^n_1\succ\Sigma^n_2\succ\dots\succ\Sigma^n_{l_n}.
\end{cases}\]

\subsection{Inter-profile axioms}
\label{subsection:inter-profile axioms}

Our main focus is on the profile, i.e., a list of possibly multiple power relations. This naturally motivates us to think of the relationships among power relations: how they interact, compensate for one another, or reinforce each other. Specifically, we will introduce three families of \textit{inter-profile} axioms: Definition \ref{axiom:CUT}, Definition \ref{axiom:CON}, and Definition \ref{axiom:UCON}.

\begin{definition}[cut and clear cut]
Let $\succsim,\succsim_1,\succsim_2\in \mathcal{R}$ with $\succsim:\Sigma_{1}\succ\Sigma_{2}\succ\dots\succ\Sigma_{l}$. We say that $(\succsim_1,\succsim_2)$ is a \textit{cut} of $\succsim$ if there exist $\hat{k}\in[l]$ and $\Gamma_1,\Gamma_2\subseteq \Sigma_{\hat{k}}$ such that 
\begin{align*}
&\succsim_1=\succsim\mid_{\Sigma_{1}\cup\dots\cup\Sigma_{\hat{k}-1}\cup\Gamma_1},\\
&\succsim_2=\succsim\mid_{\Gamma_{2}\cup\Sigma_{\hat{k}+1}\cup\dots\cup\Sigma_{l}},\\
&\Sigma_1\cup\Gamma_2=\Sigma_{\hat{k}},\text{ and}\\
&\Gamma_1\cap\Gamma_2\neq\emptyset.
\end{align*}
If $\Gamma_1=\Gamma_2=\Sigma_{\hat{k}}$, we say that $(\succsim_1,\succsim_2)$ is a \textit{clear cut} of $\succsim$. 
\end{definition}

\begin{definition}[CUT and WCUT]
\label{axiom:CUT}
An SRS $R$ satisfies \textit{cut equivalence} (CUT) [resp. \textit{weak cut equivalence} (WCUT)] if for any $n\in\mathbb{N}_{+}$, $\pi=\mset{\succsim_1,\succsim_2\dots,\succsim_n}\in\Pi$, $i,j\in[n]$, and $\succsim\in\mathcal{R}$ such that $(\succsim_i,\succsim_j)$ is a cut [resp. clear cut] of $\succsim$, we have that $R_{\pi}=R_{\pi'}$, where $\pi'$ is the profile obtained from $\pi$ by replacing two power relations $\succsim_i$ and $\succsim_{j}$ with $\succsim$. 
\end{definition}

The first family consists of CUT and its weaker version WCUT. Their main idea is simple. Suppose that we ``cut'' some power relation $\succsim$ by $\Gamma$ to obtain $\succsim_1$ (a sub-relation including $\Gamma$ and coalitions indifferent to or better than $\Gamma$) and $\succsim_2$ (a sub-relation including $\Gamma$ and coalitions indifferent to or worse than $\Gamma$). Then, CUT demands that the replacement of $\succsim$ with the pair $\succsim_1$ and $\succsim_2$ would not change the social ranking. WCUT considers the case when the worst equivalence class of $\succsim_1$ coincides with the best equivalence class of $\succsim_2$. 

To put it differently, CUT (WCUT) says that if we find two power relations $\succsim_1:\Sigma^1_1\succ\dots\succ\Sigma^1_{l_1}$ and $\succsim_2:\Sigma^2_1\succ\dots\succ\Sigma^2_{l_2}$ in a profile $\pi$ such that $\succsim_1$'s worst equivalence class $\Sigma^1_{l_1}$ and $\succsim_2$'s best equivalence class $\Sigma^2_1$ share a nonempty intersection and that is all they share, i.e., $(\bigcup_{i\leq l_1-1}\Sigma^1_{i})\cap (\bigcup_{i\geq 2}\Sigma^2_{i})=\emptyset$\footnote{This condition is necessary: if some $S$ appears both in $\Sigma^1_1,\dots,\Sigma^1_{l_1-1}$ and $\Sigma^2_2,\dots,\Sigma^2_{l_2}$, then we cannot connect $\succsim_1$ and $\succsim_2$ directly (notice that CUT [WCUT] demands that the connection must belong to $\mathcal{R}$).}, then we can unify them into one power relation with $\Sigma^1_{l_1}\cap \Sigma^2_1(\neq \emptyset)$ as a ``coupler'' between two railway cars.  

We would like to note that such a combination is typical in sports tournaments. For instance, the Grand Slam tennis tournaments usually last for weeks and consist of qualifying rounds and main draws. Technically speaking, we can distinguish between results on one day in a qualifying round and those on another day in the main draw, because the matches are conducted individually with different weather conditions and match rules. Nevertheless, we usually infer that if a doubles team $\{A,B\}$ beats $\{C,D\}$ in the main draw, and $\{C,D\}$ beats $\{E,F\}$ in the qualifying round, then we are inclined to rank $\{A,B\}$, $\{C,D\}$, and $\{E,F\}$ in this order. Our (W)CUT demands that such a combination of two power relations with some reference point would not affect the resulting social ranking.

\begin{definition}[Merging sum of power relations and profiles]
\label{definition:merging sum of profiles}
\begin{enumerate}
\item For $\mathcal{D}_1,\mathcal{D}_2\subseteq \mathcal{X}$ with $\lvert \mathcal{D}_1\rvert,\lvert \mathcal{D}_2\rvert <\infty$, $\succsim_1\in\mathcal{R}(\mathcal{D}_1)$, $\succsim_2\in\mathcal{R}(\mathcal{D}_2)$ such that $\mathcal{D}_{1}\cap\mathcal{D}_{2}=\emptyset$, $\succsim\in\mathcal{R}(\mathcal{D})$ is called a \textit{merging sum}\footnote{In \cite{suzukiConsistentSocialRanking2024,suzukiConsistenciesSocialRanking2026}, this is simply called a sum. But we call it a merging sum in order to distinguish it from the sum of profiles.} of $\succsim_1$ and $\succsim_2$ if $\mathcal{D}=\mathcal{D}_{1}\cup\mathcal{D}_{2}$, $\succsim\mid_{\mathcal{D}_1}=\succsim_1$, and $\succsim\mid_{\mathcal{D}_2}=\succsim_2$. The set of all merging sums of $\succsim_1$ and $\succsim_2$ is denoted as $\succsim_1\oplus\succsim_2$. 
\item For $\pi_1,\pi_2\in\Pi$ such that $\lvert\pi_1\rvert=\lvert \pi_2\rvert(=n)$ and $\pi_1=\mset{\succsim_1,\dots,\succsim_n}$ and $\pi_2=\mset{\succsim_1',\dots,\succsim_n'}$, $\pi=\mset{\sqsupseteq_1,\dots,\sqsupseteq_n}\in\Pi$ is called a \textit{merging sum} of $\pi_1$ and $\pi_2$ if there exists a bijection $\rho:[n]\rightarrow [n]$ such that $\sqsupseteq_i\in\succsim_i\oplus\succsim'_{\rho(i)}$ for all $i\in[n]$. The set of all merging sums of $\pi_1$ and $\pi_2$ is denoted as $\pi_1\oplus \pi_2$.  
\end{enumerate}
\end{definition}

\begin{definition}[CON and QCON]
\label{axiom:CON}
An SRS $R$ satisfies 
\begin{itemize}
\item \textit{consistency} (CON) if for any $\pi_1,\pi_2\in\Pi$ such that $\lvert \pi_1\rvert=\lvert \pi_2\rvert$ and for any $\pi\in\pi_1\oplus\pi_2$, we have that for any $x,y\in X$, (i) if $xR_{\pi_1}y$ and $xR_{\pi_2}y$, then $xR_{\pi}y$, and (ii) if $xP_{\pi_1}y$ and $xR_{\pi_2}y$, then $xR_{\pi}y$. 
\item \textit{quasi-consistency} (QCON) if for any $\pi_1,\pi_2\in\Pi$ such that $\lvert \pi_1\rvert=\lvert\pi_2\rvert$ and for any $\pi\in\pi_1\oplus\pi_2$, we have that for any $x,y\in X$, if $xP_{\pi_1}y$ and $xP_{\pi_2}y$, then $xP_{\pi}y$.
\end{itemize}
\end{definition}

Our CON is a direct extension of what is called consistency in the theory of social ranking \cite{suzukiConsistentSocialRanking2024,suzukiConsistenciesSocialRanking2026}. The original consistency says that if we partition a power relation into two disjoint ones (e.g., partitioning $\succsim^E$ into $\succsim^{E'}$ and $\succsim^{E''}$ in the discussion following Example \ref{example:researchers2}), then the social rankings $R_{\succsim^E}$, $R_{\succsim^{E'}}$, and $R_{\succsim^{E''}}$ should be coherent. We extend it so that it can be applied to a profile of power relations. QCON is a weaker version that only demands that the asymmetric part $P$ should be coherent. The relationship between QCON and CON is analogous to that between quasi-transitivity and transitivity. QCON is newly proposed in this paper in the theory of social ranking. 

\begin{definition}[UCON and QUCON]
\label{axiom:UCON}
An SRS $R$ satisfies 
\begin{itemize}
\item \textit{union consistency} (UCON) if for any $\pi_1,\pi_2\in\Pi$, we have that for any $x,y\in X$, (i) if $xR_{\pi_1}y$ and $xR_{\pi_2}y$, then $xR_{\pi_1+\pi_2}y$, and (ii) if $xP_{\pi_1}y$ and $xR_{\pi_2}y$, then $xP_{\pi_1+\pi_2}y$.
\item \textit{quasi-union consistency} (QUCON) if for any $\pi_1,\pi_2\in\Pi$, we have that for any $x,y\in X$, if $xP_{\pi_1}y$ and $xP_{\pi_2}y$, then $xP_{\pi_1+\pi_2}y$.
\end{itemize}
\end{definition}

In voting theory, consistency is a well-known axiom that demands that the social choice in two societies should be coherent with the social choice in the combined society. UCON is an analog of the axiom. Notice that in the study of social ranking, the concept of such a combination, or the sum of profiles, should arise only when the model is extended to allow profiles of power relations. In fact, we cannot find a corresponding axiom in the existing literature on social ranking.

To sum up, we have introduced three families of inter-profile axioms. It is worth noting that the first family (CUT and WCUT) is \textit{dichotomous} in that it concerns the relationship between $R_{\pi}$ and $R_{\pi'}$; but the second and third families (CON and QCON; UCON and QUCON) are \textit{trichotomous} in that they concern the relationship between $R_{\pi}$, $R_{\pi_1}$, and $R_{\pi_2}$. The logical consequences of (the combination of) these axioms will be examined later. 

\subsection{Complementary axioms}
\label{subsection:complementary axioms}

Lex-cel seems to occupy a central position in the study of SRSs. Ever since it was first proposed and characterized by \cite{bernardiRankingObjectsPreference2019c} and refined in \cite{algabaLexicographicSolutionsCoalitional2021,koniecznySelectingMostRelevant2022}, Lex-cel has been characterized/analyzed from a number of perspectives (manipulability \cite{alloucheSocialRankingManipulability2020b}; sabotage-proofness \cite{suzukiSabotageproofSocialRanking2025}; desirability \cite{aleandriDesirabilitySocialRanking2025}; consistency \cite{suzukiConsistentSocialRanking2024,suzukiConsistenciesSocialRanking2026}; independences \cite{suzukiIndependenceAxiomsSocial2025}). In most of the literature, Lex-cel is defined as a function $\mathcal{R}\rightarrow \mathcal{B}(X)$, i.e., it is a mapping from a single power relation to a complete binary relation on individuals. Below, we will first define Lex-cel as such a function, and then focus on SRSs that are considered extensions of Lex-cel. 

For $\succsim\in \mathcal{R}$ with $\succsim:\Sigma_1\succ\dots\succ\Sigma_l$ and $z\in X$, we denote the \textit{appearance vector} $\theta_{\succsim}(z)$ as $\theta_\succsim(z):=(z_1,z_2,\dots,z_l)$, where $z_k:=\lvert \{S\in \Sigma_k\mid z\in S\}\rvert$. That is, the $k$-th coordinate of $\theta_\succsim(z)$ represents how many coalitions in $\Sigma_k$ contain $z$. For two vectors of integers $a=(a_1,a_2,\dots,a_m)$ and $b=(b_1,b_2,\dots,b_m)$ with the same number of coordinates, we write $a\geq^L b$ if either (i) $a=b$ or (ii) there exists $\hat{k}\in[m]$ such that $a_k=b_k$ for all $k<\hat{k}$ and $a_{\hat{k}}>b_{\hat{k}}$. Furthermore, we write $a>^{L}b$ if $a\geq^L b$ and $\neg (b\geq^L a)$.

\begin{definition}
We define Lex-cel $L:\mathcal{R}\rightarrow \mathcal{B}(X)$ as follows. For any $\succsim\in \mathcal{R}$ and $x,y\in X$, $xL(\succsim)y\Leftrightarrow \theta_{\succsim}(x)\geq^L\theta_{\succsim}(y)$.
\end{definition}

By analogy with an SRS, we write $xL_\succsim y$ instead of $xL(\succsim)y$. Furthermore, the asymmetric (resp. symmetric) part of $L_{\succsim}$ is denoted as $P^L_{\succsim}$ (resp. $I^L_{\succsim}$). Notice that the domain of Lex-cel is usually $\mathcal{R}$ (not $\Pi$) in the literature. Hence, we define it accordingly. To reasonably extend it to the broader domain $\Pi$ is one of the main objectives of this paper. 

\begin{definition}
\label{axiom:Lex-cel extension}
An SRS $R$ satisfies \textit{Lex-cel extension} (LE) if $R_{\succsim}=L_{\succsim}$ holds for all $\succsim\in \mathcal{R}$.
\end{definition}

With this definition, our task can be rephrased as finding a reasonable SRS satisfying LE. Our approach is to focus on some essential axioms of Lex-cel. In the axiomatic literature on Lex-cel, many studies use coalitional anonymity (or its weaker version) \cite{bernardiRankingObjectsPreference2019c,algabaLexicographicSolutionsCoalitional2021,suzukiConsistentSocialRanking2024,aleandriDesirabilitySocialRankings2025,suzukiSabotageproofSocialRanking2025,suzukiConsistenciesSocialRanking2026}, and neutrality \cite{bernardiRankingObjectsPreference2019c,algabaLexicographicSolutionsCoalitional2021,suzukiConsistentSocialRanking2024,suzukiConsistenciesSocialRanking2026}. Furthermore, unlike CP majority, it is well known that Lex-cel yields transitive binary relations for any input.\footnote{It is well known that CP majority can yield non-transitive binary relations \cite{haretCeterisParibusMajority2018,suzukiUnifiedUnderstandingMajority2026a}.} We will introduce these axioms in our setting. 

\paragraph{Neutrality}
For a set $A$, let $\mathfrak{S}(A)$ denote the set of all permutations on $A$ (i.e., $\mathfrak{S}(A):=\{\sigma:A\rightarrow A\mid \sigma\text{ is a bijection}\})$). With a little abuse of notation, for $\sigma\in \mathfrak{S}(A)$ and $B\subseteq A$, we write $\sigma(B):=\{\sigma(b)\mid b\in B\}$. 

For $\sigma\in \mathfrak{S}(X)$ and $\succsim\in \mathcal{R}$, let $\succsim^\sigma\in \mathcal{R}$ be defined as follows: for any $S,T\in \mathcal{X}$, let $S\succsim T\Leftrightarrow\sigma(S)\succsim^{\sigma}\sigma(T)$. For $\pi\in \Pi$ with $\pi=\mset{\succsim_1,\dots,\succsim_n}$, let $\pi^{\sigma}:=\mset{\succsim_1^\sigma,\dots,\succsim_n^\sigma}$. In other words, $\succsim^{\sigma}$ (resp. $\pi^\sigma$) is a power relation (resp. profile) obtained from $\succsim$ (resp. $\pi$) by changing the names of individuals according to the permutation $\sigma$. 

\begin{definition}
An SRS $R$ satisfies \textit{neutrality} (NT) if for any $\pi\in \Pi$, $x,y\in X$, and $\sigma\in \mathfrak{S}(X)$, we have 
\[xR_{\pi}y\Leftrightarrow \sigma(x)R_{\pi^{\sigma}}\sigma(y).\]
\end{definition}

NT corresponds to neutrality in the literature. It is a natural axiom demanding that the names of the individuals do not matter: if we change the names of the individuals according to the permutation $\sigma$, then the social ranking should change accordingly.

\paragraph{Coalitional anonymity}
Coalitional anonymity, or its variant, is found in almost all existing characterizations of Lex-cel. According to \cite{bernardiRankingObjectsPreference2019c}, coalitional anonymity is a mixture of two principles: in determining the social ranking of $x,y\in X$, 
\begin{enumerate}[label=(\roman*)]
\item for discriminating coalitions (i.e., coalitions including exactly one of $x$ or $y$), it does not matter which sets contain $x$ and $y$, and so replacing the positions of $x$'s coalitions ($y$'s coalitions) does not change the social ranking, and 
\item the positions of nondiscriminating coalitions (i.e., coalitions including both $x$ and $y$, or coalitions including neither $x$ nor $y$) do not matter. 
\end{enumerate}
The former condition is often referred to as weak coalitional anonymity \cite{algabaLexicographicSolutionsCoalitional2021,suzukiConsistentSocialRanking2024}.

In the following, we will extend the two principles to our model. 

For $\tau\in \mathfrak{S}(\mathcal{X})$ and $S\subseteq X$, we say that $\tau$ is \textit{$S$-invariant} if for any $T\in \mathcal{X}$, we have $T\cap S=\tau(T)\cap S$. 

For $\tau\in \mathfrak{S}(\mathcal{X})$ and $\succsim\in \mathcal{R}$, we define $\succsim^\tau\in \mathcal{R}$\footnote{For conciseness of notation, for both $\sigma\in \mathfrak{S}(X)$ (i.e., a permutation of individuals) and $\tau\in \mathfrak{S}(\mathcal{X})$ (i.e., a permutation of coalitions), we denote the corresponding transformation of $\succsim$ using superscripts as $\succsim^\sigma$ and $\succsim^\tau$. But we believe that this does not cause any confusion, as long as the permutation is explicitly defined.} as follows: for any $S,T\in \mathcal{X}$, we have 
\[S\succsim T\Leftrightarrow \tau(S)\succsim^{\tau}\tau(T).\]
For $\tau\in \mathfrak{S}(\mathcal{X})$ and $\pi=\mset{\succsim_1,\dots,\succsim_n}\in \Pi$, let $\pi^{\tau}:=\mset{\succsim_1^\tau,\dots,\succsim_n^\tau}$. 

\begin{definition}
\label{axiom:weak coalitional anonymity}
An SRS $R$ satisfies \textit{weak coalitional anonymity} (WCA) if for any $\pi\in \Pi$, $x,y\in X$, and $\tau\in \mathfrak{S}(\mathcal{X})$ that is $\{x,y\}$-invariant, we have that 
\[R_{\pi}\mid_{\{x,y\}}=R_{\pi^\tau}\mid_{\{x,y\}}.\]
\end{definition}

For any $S\in \mathcal{X}$ and $x,y\in X$, we say that $S$ is $\{x,y\}$-\textit{nondiscriminating} if $S\cap \{x,y\}$ is either $\{x,y\}$ or $\emptyset$. 

\begin{definition}
\label{axiom:independence of non-discriminating sets}
An SRS $R$ satisfies \textit{independence of nondiscriminating sets} (INDS) if for any $\pi,\pi'\in\Pi$ with $\pi=\mset{\succsim_1,\dots,\succsim_n}$ and $\pi'=\mset{\succsim'_1,\dots,\succsim'_n}$, if $\succsim_i'$ is obtained from $\succsim_i$ by shifting the positions of $\{x,y\}$-nondiscriminating sets\footnote{Technically speaking, the condition means that $\succsim_i,\succsim_i'\in \mathcal{R}(\mathcal{D}_i)$ (they have the same support) and $\succsim_i\mid_{\mathcal{E}}=\succsim_i'\mid_{\mathcal{E}}$, where $\mathcal{E}:=\{S\in \mathcal{X}\mid S\text{ is }\{x,y\}\text{-nondiscriminating}\}$.} for all $i\in[n]$, then we have $R_{\pi}\mid_{\{x,y\}}=R_{\pi'}\mid_{\{x,y\}}$.
\end{definition}

Let $x,y\in X$. We say that $\succsim:\Sigma_{1}\succ\dots\succ\Sigma_{l}(\in \mathcal{R})$ is \textit{symmetric with respect to $x$ and $y$} if $\succsim=\succsim^{\sigma}$, where $\sigma$ is a permutation of $X$ that transposes $x$ and $y$ (i.e., $\sigma\in \mathfrak{S}(X)$, $\sigma(x)=y$, $\sigma(y)=x$, and $\sigma(z)=z$ for all $z\neq x,y$). 

\begin{definition}
\label{axiom:independence of symmetric power relations}
An SRS $R$ satisfies \textit{independence of symmetric power relations} (ISPR) if for any $\pi\in \Pi$, $x,y\in X$, and $\succsim\in \mathcal{R}$ that is symmetric with respect to $x$ and $y$, we have that 
\[R_{\pi}\mid_{\{x,y\}}=R_{\pi+\mset{\succsim}}\mid_{\{x,y\}}.\]
\end{definition}

The first axiom, WCA, is a straightforward extension of weak coalitional anonymity in the literature. In determining the social ranking between $x$ and $y$, WCA demands that their colleagues do not play essential roles: even if we replace some set $S$ with $S'$, as long as $S\cap \{x,y\}=S'\cap \{x,y\}$ holds, the social ranking would not change. 

The next axiom, INDS, is also a straightforward extension of principle (ii) of the original coalitional anonymity. INDS says that the shifting of $\{x,y\}$-nondiscriminating sets would not affect the social ranking between $x$ and $y$.

The last axiom, ISPR, is inspired by the principle (ii) of coalitional anonymity. ISPR demands that the social ranking between $x$ and $y$ should not depend on power relations that are symmetric with respect to $x$ and $y$: even if we drop such symmetric ones, the social ranking would not change. Notably, ISPR allows us to focus only on the essential difference between $x$ and $y$. In comparing two researchers, for instance, there are infinitely many characteristics that ``could be'' used to compare them, such as race, sex, citations, and educational history. But many of them, such as race or sex, usually have no discriminating power in the sense that the candidates are completely indifferent according to the criteria. A practical benefit of ISPR is that we can disregard such nonessential power relations.

\paragraph{Transitivity, quasi-transitivity, and triple-acyclicity}

\begin{definition}
An SRS $R$ satisfies 
\begin{itemize}
\item \textit{transitivity} (TR) if $R_{\succsim}$ is transitive for all $\succsim\in \mathcal{R}$;
\item \textit{quasi-transitivity} (QT) if $R_{\succsim}$ is quasi-transitive for all $\succsim\in \mathcal{R}$ (i.e., for any $x,y,z\in X$, we have $[(xP_{\succsim}y\land yP_{\succsim}z)\Rightarrow xP_{\succsim}z]$);
\item \textit{triple-acyclicity} (TA) if $R_{\succsim}$ is triple-acyclic for all $\succsim\in \mathcal{R}$ (i.e., for any $x,y,z\in X$, we have $[(xP_{\succsim}y \land yP_{\succsim}z)\Rightarrow \neg (zP_{\succsim}x)]$).
\end{itemize}
\end{definition}

TR, QT, and TA demand that the social ranking $R_{\succsim}$ should satisfy the corresponding rationality conditions. It is easy to verify that $L_{\succsim}$ is transitive for all $\succsim\in \mathcal{R}$ (see, e.g., \cite{bernardiRankingObjectsPreference2019c}). 

\section{Impossibilities concerning multiple inter-profile axioms}
\label{section:multiple inter-profile axioms}

We are ready to present our main results. This section first examines the logical consequences of the combination of inter-profile axioms introduced in Section \ref{subsection:inter-profile axioms}. All proofs of the theorems are in Appendix \ref{appendix:proof of theorems}.

We define an SRS $R^{U}$ (\textit{unanimous rule}) as follows. For any $\pi\in \Pi$, $R^U_\pi:=X\times X$ (i.e., $xI^U_{\pi}y$ for all $x,y\in X$). This is an extension of the unanimous rule in the literature \cite{morettiAxiomaticAlgorithmicPerspectives2017}. Notice that the unanimous rule is not an interesting SRS, as it constantly judges any pair of individuals as indifferent. 

\begin{theorem}
\label{theorem:CUT and CON}
An SRS satisfies CUT, CON, and NT if and only if it is the unanimous rule $R^{U}$. 
\end{theorem}

\begin{theorem}
\label{theorem:CUT and UCON}
An SRS satisfies CUT, UCON, and NT if and only if it is the unanimous rule $R^{U}$. 
\end{theorem}

\begin{theorem}
\label{theorem:CON and UCON}
No SRS satisfies CON, UCON, and LE simultaneously. 
\end{theorem}

Theorems \ref{theorem:CUT and CON}, \ref{theorem:CUT and UCON}, and \ref{theorem:CON and UCON} show basic impossibilities concerning the pairs of the three inter-profile axioms. Under NT, which should be one of the most natural axioms in social choice theory, we cannot meet CUT and CON/UCON simultaneously (Theorems \ref{theorem:CUT and CON} and \ref{theorem:CUT and UCON}). Furthermore, the pair of CON and UCON is not promising either, as the pair is incompatible with LE (Theorem \ref{theorem:CON and UCON}). 

These impossibilities can be partly attributed to the different philosophies of the three families of inter-profile axioms. To briefly rephrase the discussion in Section \ref{subsection:inter-profile axioms}, each of CUT, CON, and UCON requires that the social ranking should behave in relation to certain operations on the profiles. As such ``operations'', CUT considers the ``connection'' of two power relations using some coalitions as the coupler; CON considers the ``merge'' of two profiles by taking the merging sum of corresponding pairs of power relations; UCON considers the ``unification'' of two profiles by simply taking the sum of multisets. The differences between these operations (connection, merge, and unification) are described as follows. 
\begin{itemize}
\item The connection can change how many times each coalition appears in the profile, while the merge and unification do not. For instance, if we compare $\pi$ and $\pi'$ in Section \ref{section:introduction}, $\{3\}$ appears twice in the former but only once in the latter (because $\{3\}$ behaved as the ``coupler''). But if we consider the merge or unification, such a change does not occur, because each coalition should be kept (they do not disappear) through the process of merge or unification. 
\item The merging operation is highly flexible. As Suzuki and Horita \cite{suzukiConsistentSocialRanking2024} also demonstrate (their Example 1), two power relations can have many merging sums. For instance, in the example of Section \ref{section:introduction}, while $\succsim^E$, $\succsim^{E'''}:\{\{1,2\}\}\succ\{\{1,4\}\}\succ\{\{1,3\}\}\succ\{\{2,3\}\}\succ\{\{3\}\}$ and $\succsim^{E''''}\{\{1,2\},\{1,3\}\}\succ\{\{2,3\}\}\succ\{\{1,4\},\{3\}\}$ are also elements of $\succsim^{E'}\oplus\succsim^{E''}$. Moreover, because of the bijection $\rho$ in Definition \ref{definition:merging sum of profiles}, the flexibility increases in our setting (i.e., we can find many ways to select the corresponding power relations whose merging sum is considered). This high flexibility is particular to CON compared with CUT or UCON.
\item In one sense, unification is the least controversial among the three, because it does not make any change to the original power relations; the unification simply adds them into one multiset. Another noteworthy property of UCON is that it disregards the repetition of the same profile. For instance, consider the following two profiles. 
\begin{align*}
\pi:\begin{cases}
\succsim_1&:\{\{x\}\}\succ\{\{y\}\}
\end{cases}\\
\pi':\begin{cases}
\succsim_1:&:\{\{x\}\}\succ\{\{y\}\}\\
\succsim_2:&:\{\{x\}\}\succ\{\{y\}\}
\end{cases}
\end{align*}
We can see that $\pi'=\pi+\pi$ ($\pi'$ is obtained from $\pi$ by repeating $\pi$ twice). In this case, UCON requires that $R_{\pi}=R_{\pi'}$. While many evaluation methods in use should satisfy this property, there can be counterexamples. For instance, consider threshold-based methods such as ``judging $x$ better than $y$ if the number of $x$-supporting power relations (indicators) is at least $t$ (threshold) greater than that of $y$.'' Such a method should be reasonable if we interpret each $\succsim_1$ as a sample/dataset. If we interpret each $\succsim_t$ as the dataset for year $t$, it would be necessary to collect a sufficiently large profile of datasets (i.e., rather than $\succsim_{2026}$ only, we need $\succsim_{2017},\dots,\succsim_{2026}$) to obtain statistically significant conclusions. 
\end{itemize}

\section{Extending Lex-cel}
\label{section:extending Lex-cel}

\subsection{Further impossibilities}
\label{section: further impossibilities}

From now on, we will consider a reasonable extension of Lex-cel. In doing so, the impossibilities in Section \ref{section:multiple inter-profile axioms} lead us to focus on exactly one of the inter-profile axioms (CUT, CON, and UCON), because any pair of the three, under some plausible axiom(s), has already been shown to lead to an impossibility. The following Theorems \ref{theorem:LE and WCUT}, \ref{theorem:LE and CON}, and \ref{theorem:LE and UCON} show some results in this direction. 

\begin{theorem}
\label{theorem:LE and WCUT}
No SRS satisfies LE, WCUT, and INDS.
\end{theorem}

\begin{theorem}
\label{theorem:LE and CON}
No SRS satisfies LE, CON, and ISPR.
\end{theorem}

Notice that ISPR is a direct consequence of LE and UCON.\footnote{This can be verified as follows. Let $\pi\in \Pi$ and $x,y\in X$. If $\succsim\in \mathcal{R}$ is symmetric with respect to $x$ and $y$, then $\theta_{\succsim}(x)=\theta_{\succsim}(y)$ clearly holds. So, LE implies that $xI_{\succsim}y$. Therefore, according to UCON, the addition of $\succsim$ to $\pi$ does not change the social ranking between $x$ and $y$.} So, Theorem \ref{theorem:LE and CON} can be considered as a corollary of Theorem \ref{theorem:CON and UCON}. 

\begin{theorem}
\label{theorem:LE and UCON}
No SRS satisfies LE, UCON, INDS, and TA.
\end{theorem}

These three theorems consider ways to extend Lex-cel in accordance with each one of the three inter-profile axioms. But the results are not promising: no matter which inter-profile axiom is applied, we cannot extend Lex-cel while preserving the basic axioms we introduced in Section \ref{subsection:complementary axioms}. These results motivate us to consider weaker versions of the axioms (as WCUT did not work in Theorem \ref{theorem:LE and WCUT}, we focus on QCON and QUCON). In fact, this guides us to a basic possibility in the next section.

\subsection{Characterization of PEL}
\label{section:characterization of PEL}

We define a new SRS $R^{PEL}$ called \textit{Pareto-extension Lex-cel} as follows. For any $\pi\in \Pi$ and $x,y\in X$, we have 
\[
xR^{PEL}_{\pi}y\Leftrightarrow \neg(\forall \succsim\in \pi:yL_{\succsim}x\text{ and }\exists \succsim\in\pi:yP^{L}_{\succsim}x).
\]
$R^{PEL}$ applies the Pareto principle to the results of Lex-cel: individual $x$ is judged better than $y$ if (i) for any $\succsim\in \pi$, $xL_{\succsim}y$, and (ii) there exists $\succsim'\in \pi$ such that $xP^L_{\succsim}y$. Notice that we name Pareto-extension Lex-cel after Sen's Pareto extension rule \cite{senQuasiTransitivityRationalChoice1969}. In fact, if we identify $L_\succsim$ (i.e., Lex-cel's outcome at a power relation) with $R_i$ (individual $i$'s preference relation), then our PEL and Sen's Pareto extension rule coincide as follows: 

\begin{table}[htbp]
\centering
\caption{PEL and Sen's Pareto extension rule: correspondence}
\label{table:correspondence PEL Pareto extension rule}
\renewcommand{\arraystretch}{1.5} 
\begin{tabular}{CC}
\toprule
PEL & Sen's Pareto extension rule \\
\midrule
$L_{\succsim}$ & $i$'s preference $R_i$ \\[0.5em]
$xP_{\pi}y \iff (\forall \succsim: xL_{\succsim}y \text{ and } \exists \succsim: xP^{L}_{\succsim}y)$ & 
$xP y \iff (\forall i: xR_i y \text{ and } \exists i: xP_i y)$ \\
\bottomrule
\end{tabular}
\end{table}

\begin{example}
\label{example:PEL}
In the profile $\pi'$ in Section \ref{section:introduction}, we have $3P^{PEL}_{\pi'}2$, $2I^{PEL}_{\pi'}1$, and $1I^{PEL}_{\pi'}3$. In this case, $R^{PEL}$ fails to be transitive.
\end{example}

It turns out that this SRS $R^{PEL}$ satisfies most of the axioms (at least in their weaker versions) as follows. 

\begin{theorem}
\label{theorem:characterization of PEL}
An SRS $R$ satisfies LE, QCON, QUCON, WCA, and ISPR if and only if it is $R^{PEL}$.
\end{theorem}

We have already pointed out that under the principle of coalitional anonymity (such as ISPR or INDS) and TA, LE is incompatible with each of the inter-profile axioms (the theorems in Section \ref{section: further impossibilities}). Compared with these results, Theorem \ref{theorem:characterization of PEL} is regarded as a possibility result because it reveals that Pareto-extension Lex-cel $R^{PEL}$ satisfies weaker versions of CON and UCON as well as some other axioms. Just as Sen's Pareto extension rule \cite{senQuasiTransitivityRationalChoice1969} suggests an escape route from Arrow's impossibility theorem \cite{arrowSocialChoiceIndividual2012} by weakening transitivity to quasi-transitivity, the present result suggests an escape route from the impossibilities in the preceding sections by replacing CON and UCON with QCON and QUCON, respectively.

Further discussion will follow in Section \ref{subsection:discussion}. 

\subsection{Discussion}
\label{subsection:discussion}
Two topics are discussed in detail. 

\paragraph{Alternative extensions of Lex-cel}
It is worth noting that the way to extend Lex-cel to our setting is not unique: there are some other extensions (i.e., SRSs satisfying LE). 

For $\pi=\mset{\succsim_1,\dots,\succsim_n}\in \Pi$ with $\succsim_i:\Sigma^i_1\succ\Sigma^i_2\succ\dots\succ\Sigma^i_{l_i}$ and $z\in X$, the vector $\bar{\theta}_{\pi}(z)$ is defined as follows: let $l^{*}:=\max_{i\in[n]}l_i$. For each $i\in[n]$ and $k\in[l^*]\setminus[l_i]$, let $\Sigma^i_k:=\emptyset$. For each $i\in[n]$ and $k\in[l^*]$, let $z^i_k:=\lvert \{S\in\Sigma^i_k\mid z\in S\}\rvert$. Then, let
\[\bar{\theta}_{\pi}(z):=(\Sigma_{i\in[n]}z^i_1,\Sigma_{i\in[n]}z^i_2,\dots,\Sigma_{i\in[n]}z^i_{l^*}).\]

\begin{definition}
\label{definition:alternative extensions of Lex-cel}
SRSs called \textit{majority Lex-cel} $R^{ML}$, $\alpha$-majority Lex-cel $R^{\alpha\text{-}ML}$ (where $1/2<\alpha\leq 1$), \textit{weak-Pareto Lex-cel} $R^{WPL}$, and \textit{absolute Lex-cel} $R^{AL}$ are defined as follows. For $\pi=\mset{\succsim_1,\dots,\succsim_n}\in \Pi$ and $x,y\in X$, let
\begin{itemize}
\item $xR^{ML}_{\pi}y\Leftrightarrow \lvert \{i\in[n]\mid xR^{L}_{\succsim}y\}\rvert \geq \lvert \{i\in[n]\mid yR^{L}_{\succsim}x\}\rvert$.
\item $xP^{\alpha\text{-}ML}y\Leftrightarrow\frac{\lvert\{i\in[n]\mid xP^{L}_{\succsim_i}y\}\rvert}{n}\geq \alpha,$ 
\item $xR^{WPL}_{\pi}y\Leftrightarrow \neg (\forall \succsim\in \pi:yP^{L}_{\succsim}x)$.
\item $xR^{AL}_{\pi}y\Leftrightarrow \bar{\theta}_{\pi}(x)\geq^L\bar{\theta}_{\pi}(y)$.
\end{itemize}
\end{definition}

\begin{example}
\label{example:ML, WPL, and AL}
In the profile $\pi$ in the Introduction, we have that $1P^{ML}_{\pi}2$, $3P^{ML}_{\pi}1$, $1P^{ML}_{\pi}4$, $3P^{ML}_{\pi}2$, $2P^{ML}_{\pi}4$, and $3P^{ML}_{\pi}4$; $1I^{WPL}_{\pi}2$, $3P^{WPL}_{\pi}1$, $1P^{WPL}_{\pi}4$, $3P^{WPL}_{\pi}2$, $2P^{WPL}_{\pi}4$, and $3P^{WPL}_{\pi}4$. Furthermore, we have $\bar{\theta}_{\pi}(1)=(1,3,2,0)$, $\bar{\theta}_{\pi}(2)=(1,2,2,0)$, $\bar{\theta}_{\pi}(3)=(3,2,1,1)$, and $\bar{\theta}_{\pi}(0,0,1,1)$, resulting in $3P^{AL}_{\pi}1P^{AL}_{\pi}2P^{AL}_{\pi}4$. In general, we have that $R^{1\text{-}ML}=R^{WPL}$.
\end{example}

The intuition behind these SRSs is as follows. $R^{ML}$, $R^{\alpha\text{-}ML}$, and $R^{WPL}$ are analogs of majority rule, supermajority rule (with threshold $\alpha$), and the weak Pareto principle in voting theory. In fact, if we identify Lex-cel's outcome $L_{\succsim}$ with a voter's preference (just as we did in Table \ref{table:correspondence PEL Pareto extension rule}), we can verify the respective correspondences. The above three SRSs, as well as Pareto-extension Lex-cel $R^{PEL}$, all apply Lex-cel to each power relation and then aggregate the outcomes using some voting procedure. In contrast, $R^{AL}$ first aggregates the appearance vector $\theta_{\succsim}(z)$ of each power relation $\succsim\in\pi$ into $\bar{\theta}_{\pi}(z)$, and then applies Lex-cel. 

Although these extensions look reasonable at first glance, they fail to satisfy some of the complementary axioms introduced so far. In fact, $R^{AL}$ fails to satisfy INDS, $R^{WPL}$ fails to satisfy ISPR, and $R^{ML}$ fails to satisfy QCON (Propositions \ref{proposition:ind LE,UCON,INDS,TA} and \ref{proposition:ind ISPR,WCA,QUCON,QCON,LE}).

Finally, it is worth reviewing these SRSs in terms of social rationality (TR, QT, and TA).

\begin{proposition}
\label{proposition:lex-cel extensions}
\begin{itemize}
\item $R^{ML}$ fails to satisfy TA.
\item For $1>\alpha>\frac{2}{3}$, $R^{\alpha\text{-}ML}$ satisfies TA, but not QT.
\item Both $R^{WPL}$ and $R^{PEL}$ satisfy QT, but not TR.
\item $R^{AL}$ satisfies TR.
\end{itemize}
\end{proposition}

\begin{proof}
The first two statements are easy to verify by analogy with majority rule, supermajority rule, weak Pareto principle, and Pareto extension rule. We demonstrate only the nontrivial part here. 

For $R^{ML}$, consider
\[\pi:\begin{cases}
\succsim_1:\{\{x\}\}\succ\{\{y\}\}\succ\{\{z\}\}\\
\succsim_2:\{\{y\}\}\succ\{\{z\}\}\succ\{\{x\}\}\\
\succsim_3:\{\{z\}\}\succ\{\{x\}\}\succ\{\{y\}\}.
\end{cases}\]
Then, $xP^{ML}_{\pi}y$, $yP^{ML}_{\pi}z$, and $zP^{ML}_{\pi}x$ hold. So, $R^{ML}$ violates TA (this is essentially equivalent to Condorcet's paradox). 

Let $\frac{2}{3}<\alpha<1$. Then, $R^{\alpha\text{-}ML}$ satisfies TA. This is shown as follows. Let $\pi=\mset{\succsim_1,\succsim_2,\dots,\succsim_n}$ and assume that $xP_{\pi}^{\alpha\text{-}ML}y$ and $yP^{\alpha\text{-}ML}_{\pi}z$. By definition, 
\[
\frac{\{i\in[n]\mid xP^{L}_{\succsim_i}y\}}{n}\geq \alpha\text{ and }
\frac{\{i\in[n]\mid yP^{L}_{\succsim_i}z\}}{n}\geq \alpha.\\
\]
Hence, for at least $2\alpha-1(>\frac{1}{3})$ elements among $\succsim_1,\succsim_2,\dots,\succsim_n$, we have both $xP^{L}_{\succsim_i}y$ and $yP^{L}_{\succsim_i}z$, which means $xP^{L}_{\succsim_i}z$ by the transitivity of Lex-cel. In other words, $zP^{L}_{\succsim_i}x$ can hold for at most $1-(2\alpha-1)(<\frac{2}{3}<\alpha)$ power relations in $\pi$. So, $xR_{\pi}^{\alpha\text{-}ML}z$ holds. This proves that $R^{\alpha\text{-}ML}$ ($\frac{2}{3}<\alpha<1$) satisfies TA. 

However, $R^{\alpha\text{-}ML}$ ($\frac{2}{3}<\alpha<1$) does not satisfy QT. This is shown as follows. As $\alpha<1$, there exist $p,q\in\mathbb{N}$ such that $\alpha<\frac{q}{p}<1$. Now, let $\pi:=\mset{\succsim_1,\succsim_2,\dots,\succsim_p}$ such that 
\begin{align*}
\succsim_i&:\{\{z\}\}\succ\{\{x\}\}\succ\{\{y\}\}\quad\text{for }1\leq i\leq p-q,\\
\succsim_i&:\{\{x\}\}\succ\{\{y\}\}\succ\{\{z\}\}\quad\text{for }p-q+1\leq i\leq q,\\
\succsim_i&:\{\{y\}\}\succ\{\{z\}\}\succ\{\{x\}\}\quad\text{for }q+1\leq i\leq p.
\end{align*}
Then, we can easily verify that $xP^{L}_{\succsim_i}y$ for $1\leq i\leq q$ and $yP^{L}_{\succsim_i}z$ for $p-q+1\leq i\leq p$. These show that $xP^{\alpha\text{-}ML}_\pi y$ and $yP^{\alpha\text{-}ML}_{\pi}z$. However, $xP^{\alpha\text{-}ML}_{\succsim_i}z$ holds only when $p-q+1\leq i\leq q$. Notice that this interval $p-q+1\leq i\leq q$ has exactly $2q-p$ integers, and $2q-p<q$ holds by our assumption that $\frac{q}{p}<1$. Hence, $zR^{\alpha\text{-}ML}_{\pi}x$ must hold. 

For $R^{WPL}$ and $R^{PEL}$, it is straightforward to verify that they satisfy QT. But they do not satisfy TR: counterexamples are already shown in Examples \ref{example:ML, WPL, and AL} and \ref{example:PEL}, respectively.

The proposition that $R^{AL}$ satisfies TR is straightforward because the social ranking obtained by Lex-cel is always transitive \cite{bernardiRankingObjectsPreference2019c,suzukiConsistenciesSocialRanking2026} (notice that $R^{AL}_{\pi}$ is obtained by simply applying $\geq ^L$ to the vectors $\bar{\theta}_{\pi}(\cdot)$).
\end{proof}

Notice that $R^{ML}$, $R^{WPL}$, $R^{PEL}$, and $R^{AL}$ all apply Lex-cel in some way. The difference is that the first three apply Lex-cel to each power relation, and then aggregate the resulting rankings using some voting procedure, whereas the last one ($R^{AL}$) aggregates the appearance vectors and applies a Lex-cel-like procedure in the final phase (this is why TR continues to hold). 

\paragraph{Arrow's impossibility}
It is interesting to examine the relationship between Arrow's impossibility theorem and our Theorem \ref{theorem:LE and UCON} in more detail. Table \ref{table:correspondence Arrow and our theorem} visualizes their relationship.\footnote{An acute reader might wonder why the results differ: our Theorem \ref{theorem:LE and UCON} does not even allow ``dictatorship''. This is because our model treats each power relation equally by definition (recall that a profile of power relations is simply a multiset: two indicators with the same power relation are considered identical).}

\begin{table}[htbp]
\centering
\caption{Arrow's impossibility and Theorem \ref{theorem:LE and UCON}: correspondence}
\label{table:correspondence Arrow and our theorem}
\renewcommand{\arraystretch}{1.4}
\newcolumntype{L}{>{\raggedright\arraybackslash}m{5cm}}

\begin{tabular}{LL}
\toprule
\hfil \textbf{Arrow's impossibility} \cite{arrowSocialChoiceIndividual2012} & 
\hfil \textbf{Our Theorem \ref{theorem:LE and UCON}} \\
\midrule
\textbf{Unrestricted domain}: all logically possible profiles are admitted & 
\textbf{The domain $\Pi$} contains all logically possible profiles of power relations \\[0.8em]

\textbf{Pareto principle}: $(\forall i: xP_i y \Rightarrow xPy)$ & 
\textbf{LE and UCON}: $(\forall \succsim\, \in \pi: xP^{L}_{\succsim}y \Rightarrow xP_\pi y)$ \\[0.8em]

\textbf{Independence of irrelevant alternatives}: positions of irrelevant alternatives do not matter & 
\textbf{INDS}: positions of nondiscriminating coalitions do not matter \\
\midrule[\lightrulewidth] 
\hfil \textbf{Results: Dictatorship} & \hfil \textbf{Results: No such SRS exists} \\
\bottomrule
\end{tabular}
\end{table}

The table provides an intuitive explanation of why Theorem \ref{theorem:LE and UCON} should hold by contrasting the models of Arrow's impossibility theorem and our setting. 

This observation partly supports the interpretation of our SRSs as an instance of the Arrovian aggregation problem, but we would like to emphasize that the study of SRSs can be more than that for several reasons. The first reason is that an SRS usually requires some operations on the power relations. As \cite{suzukiIndependenceAxiomsSocial2025} argues, it is often too computationally/cognitively demanding to rank all the logically possible coalitions (as $n$ individuals can form $2^n$ coalitions). This supports the necessity of connection or (merging) sum of the power relations, which is not common in the classical aggregation problem. The second reason lies in the interpretation of the Pareto principle. In the classical Arrovian model, in which aggregation of preferences over the set of alternatives matters, the meaning of the Pareto principle is rather straightforward, because whether $x$ ``is better than'' $y$ is explicit in someone's preferences. However, in the social ranking model, to say $x$ is better than $y$ under some power relation $\succsim$ can be ambiguous, because each individual ($x,y,\dots$) can appear several times in a variety of coalitions, e.g., 
\[\succsim:\{\{1,3\}\}\succ\{\{2,3\},\{2,4\},\{2,3,4\}\}\succ\{\{1,4\},\{1,3,4\}\}\succ\{\{2\}\}\succ\{\{1\}\}\]
(both $1$ and $2$ appear four times). Lex-cel focuses on the appearances in high classes, and so it says that $1P^L_{\succsim}2$. But if we focus on the ceteris paribus comparison instead, $2$ should be judged better than $1$, because $2$ wins more times than $1$ in ceteris paribus comparisons: comparing $S\cup \{2\}$ and $S\cup \{1\}$ for $S\subseteq X\setminus\{1,2\}$, we can observe that $\{2\}\succ\{1\}$, $\{2,4\}\succ\{1,4\}$, $\{2,3,4\}\succ\{1,3,4\}$ (these three comparisons support $2$), and $\{1,3\}\succ \{2,3\}$ (this comparison supports $1$). Such a judgment is called CP majority in the literature \cite{haretCeterisParibusMajority2018}. 

This observation suggests another interesting question: what happens to Theorem \ref{theorem:LE and UCON} if we focus on CP majority, instead of Lex-cel, as the counterpart of the Pareto principle in Arrow's impossibility theorem? A study closely related to this question is \cite{suzukiUnifiedUnderstandingMajority2026a}. They determine the necessary and sufficient conditions for CP majority (and its variants) to be transitive in a slightly different model\footnote{According to their opinion aggregation model, which originates in \cite{barberaOpinionAggregationBorda2023}, each ground (power relation in our context) is not distinguished: the model only considers the number of grounds supporting some coalitions.}.

\section{Conclusion}
\label{section:conclusion}

We have extended the social ranking model to allow multiple indicators (criteria, reviewers, data sources, etc.), and axiomatically analyzed reasonable SRSs in the extended model. The results are summarized in Table \ref{table:Summary of the results}\footnote{Axioms in brackets, NT and QT, are not used in the characterization, but they are satisfied by PEL.}. While we showed a series of negative results (Theorems \ref{theorem:CUT and CON}, \ref{theorem:CUT and UCON}, \ref{theorem:CON and UCON}, \ref{theorem:LE and WCUT}, \ref{theorem:LE and CON}, \ref{theorem:LE and UCON}), these can be classified into several categories: the incompatibilities between inter-profile axioms (Theorems \ref{theorem:CUT and CON}, \ref{theorem:CUT and UCON}, and \ref{theorem:CON and UCON}), and the incompatibilities between LE and each inter-profile axiom (Theorems \ref{theorem:LE and WCUT}, \ref{theorem:LE and CON}, \ref{theorem:LE and UCON}). By analogy with Sen's Pareto extension rule as an escape route from Arrow's impossibility theorem, we show that Pareto-extension Lex-cel $R^{PEL}$ satisfies several inter-profile axioms as well as some main properties of Lex-cel (Theorem \ref{theorem:characterization of PEL}). Further investigation of SRSs under multiple indicators, such as the extensions of other SRSs studied so far, could be an interesting line of future research.

\begin{table}[htbp]
\centering
\caption{Summary of the results} 
\label{table:Summary of the results} 
\renewcommand{\arraystretch}{1.3}
\setlength{\tabcolsep}{5pt}

\begin{tabularx}{\textwidth}{l*{8}{>{\centering\arraybackslash}X}}
\toprule
 & \shortstack[c]{CUT/WC\\UT}
 & \shortstack[c]{CON/QC\\ON}
 & \shortstack[c]{UCON/Q\\UCON}
 & LE
 & NT
 & \shortstack[c]{WCA/IN\\DS/ISPR}
 & TA
 & Result \\
\midrule
Thm 1 & CUT & CON & & & NT & & & Unanimous \\
Thm 2 & CUT & & UCON & & NT & & & Unanimous \\
Thm 3 & & CON & UCON & LE & & & & None \\
Thm 4 & WCUT & & & LE & & INDS & & None \\
Thm 5 & & CON & & LE & & ISPR & & None \\
Thm 6 & & & UCON & LE & & INDS & TA & None \\
Thm 7 & & QCON & QUCON & LE & (NT)
      & WCA,ISPR & (QT) & PEL \\
\bottomrule
\end{tabularx}
\end{table}

\section*{Funding}

This work was supported by JSPS KAKENHI Grant Numbers JP26K04808 and JP25K01302.

\section*{Declaration of competing interests}

The author declares that they have no known competing financial interests or personal relationships that could have appeared to influence the work reported in this paper.

\section*{Declaration of generative AI and AI-assisted technologies in the manuscript preparation process}
During the preparation of this work, the author used ChatGPT (OpenAI) to search for academic English expressions and usage examples. The author takes full responsibility for the content of the published article.

\begin{appendices}

\section{Proofs of the theorems}
\label{appendix:proof of theorems}

\begin{proof}[Proof of Theorem \ref{theorem:CUT and CON}]
The `if' part is straightforward. We will prove the `only if' part. Let $R$ be an SRS satisfying CUT, CON, and NT. 

The proof proceeds in three steps.

\textbf{Step 1.} We will prove the following claim. 

\begin{claim}
\label{claim:single-class unanimous}
For any single-class power relation $\succsim\in\mathcal{R}$, we have $R_{\succsim}=R^{U}_{\succsim}$.
\end{claim}

Let $\succsim\in\mathcal{R}$ be any single-class power relation such that $\succsim:\Sigma$. Fix any distinct $x,y\in X$. Suppose that there exists $S\subseteq X\setminus\{x,y\}$ such that $S\cup\{x\}\in \Sigma$ and $S\cup\{y\}\notin \Sigma$. Take any distinct $z,w\in X\setminus\{x,y\}$ that do not appear in $\pi$ (we say that \textit{$z$ appears in $\pi$} if $\pi$ contains a power relation $\succsim:\Sigma_1\succ\dots\succ\Sigma_l$ such that at least one coalition $S\in \Sigma_1\cup\dots\cup\Sigma_l$ contains $z$).  

Let us define $\pi_1,\dots,\pi_8$ and $\succsim_{1},\dots,\succsim_{10}$ as follows: 
\begin{align*}
&\pi:\begin{cases}
\succsim:\Sigma
\end{cases}\\
&\pi_1:\begin{cases}
\succsim_{1}:\{\{z\},\{w\}\}
\end{cases}\\
&\pi_2:\begin{cases}
\succsim_{2}:\Sigma\cup\{\{z\},\{w\}\}\end{cases}\\
&\pi_3:\begin{cases}
\succsim_3:\Sigma\cup\{\{z\}\}\\
\succsim_1
\end{cases}\\
&\pi_4:\begin{cases}
\succsim_4:\{\{z,w\}\},\\
\succsim_5:\{S\cup\{x\},S\cup\{y\}\}
\end{cases}\\
&\pi_5:\begin{cases}
\succsim_6:\Sigma\cup\{\{z\},\{z,w\}\}\\
\succsim_7:\{\{z\},\{w\},S\cup\{x\},S\cup\{y\}\}\\
\end{cases}\\
&\pi_6:\begin{cases}
\succsim_8:\Sigma\cup\{\{z\},\{w\},\{z,w\},S\cup\{y\}\}
\end{cases}\\
&\pi_7:\begin{cases}
\succsim_9:\{\{z\},\{w\},\{z,w\}\}
\end{cases}\\
&\pi_8:\begin{cases}
\succsim_{10}:\Sigma\cup\{S\cup\{y\}\}
\end{cases}
\end{align*}

Then, we can confirm that $R_{\pi}\mid_{\{x,y\}}=R_{\pi_8}\mid_{\{x,y\}}$ as follows. First, $xI_{\pi_1}y$ holds by NT. Using CON, we have $R_{\pi}\mid_{\{x,y\}}=R_{\pi_2}\mid_{\{x,y\}}$. CUT implies $R_{\pi_2}\mid_{\{x,y\}}=R_{\pi_3}\mid_{\{x,y\}}$ as well. NT implies $xI_{\pi_4}y$. Notice that $\succsim_6\in \succsim_3\oplus\succsim_4$ and $\succsim_7\in\succsim_1\oplus\succsim_5$. So, $\pi_5\in\pi_3\oplus\pi_4$. Thus, CON implies $R_{\pi_3}\mid_{\{x,y\}}=R_{\pi_5}\mid_{\{x,y\}}$. With CUT, we have $R_{\pi_5}\mid_{\{x,y\}}=R_{\pi_6}\mid_{\{x,y\}}$ (recall that $S\cup\{x\}\in \Sigma$). We have $xI_{\pi_7}y$ by NT. Notice that $\pi_6\in\pi_7\oplus\pi_8$. Hence, $R_{\pi_8}\mid_{\{x,y\}}=R_{\pi_6}\mid_{\{x,y\}}$. To sum up, we can conclude that $R_{\pi}\mid_{\{x,y\}}=R_{\pi_{8}}\mid_{\{x,y\}}$.

According to the above argument, by adding some elements to $\Sigma$, we can find $\succsim':\Sigma'$, where $\Sigma'\supseteq \Sigma$, such that $R_{\pi}\mid_{\{x,y\}}=R_{\pi'}\mid_{\{x,y\}}$ and $S\cup\{x\}\in \Sigma'\Leftrightarrow S\cup \{y\}\in \Sigma'$ for all $S\subseteq X\setminus \{x,y\}$ (i.e., $\succsim'$ is symmetric with respect to $x$ and $y$). Hence, NT says that $xI_{\pi'}y$, and so we obtain that $xI_{\pi}y$. This proves Claim \ref{claim:single-class unanimous}. 

\vspace{0.5em}

\textbf{Step 2.} We will prove that for any $\succsim\in \mathcal{R}$, $R_{\succsim}=R^{U}_{\succsim}$. To prove this, take any power relation $\succsim\in \mathcal{R}$. Let $\succsim:\Sigma_{1}\succ\Sigma_{2}\succ\dots\succ\Sigma_{l}$. For each $k\in[l]$, let $\succsim_k:\Sigma_{k}$. By Claim \ref{claim:single-class unanimous}, we have $xI_{\succsim_{k}}y$. Since $\succsim\in((\succsim_1\oplus\succsim_2)\oplus\succsim_3)\dots\oplus\succsim_l$, CON implies $xI_{\succsim}y$.

\vspace{0.5em}

\textbf{Step 3.} We will prove that for any $\pi\in \Pi$, $R_{\pi}=R^{U}_{\pi}$. The proof proceeds by induction on $\lvert \pi\rvert$. The case of $\lvert\pi\rvert=1$ was proved in Step 2. Assume that $R_\pi=R_\pi^U$ for all $\pi$ with $\lvert \pi\rvert\leq n$. Take any $\pi\in\Pi$ with $\lvert \pi\rvert=n+1$. Label the elements of $\pi=\mset{\succsim_1,\succsim_2,\dots,\succsim_n}$. For each $k\in[n]$, let $\succsim_i:\Sigma^i_1\succ\Sigma^i_2\succ\dots\succ\Sigma^i_{l_i}$. Fix any $z\in X$ that does not appear in $\pi$. 

Let $\pi_1:=\mset{\succsim'_1,\dots,\succsim'_n}$, where 
\begin{align*}
\succsim_i'&:\Sigma^i_1\succ\dots\succ\Sigma^i_{l_i}\succ\{\{z\}\}\quad 1\leq i\leq n-1,\\
\succsim_n'&:\Sigma^n_1\succ\dots\succ\Sigma^n_{l_n}\succ\{\{z\}\}\succ\Sigma^{n+1}_1\succ\dots\succ\Sigma^{n+1}_{l_{n+1}}.
\end{align*}
By the induction hypothesis, we have $xI_{\pi_1}y$.

Let $\pi_2:=\mset{\succsim''_1,\dots,\succsim''_{n+1}}$, where
\begin{align*}
\succsim''_{i}&:=\succsim_i'\quad 1\leq i\leq n-1,\\
\succsim''_n&:\Sigma^n_1\succ\dots\succ\Sigma^n_{l_{n}}\succ\{\{z\}\},\\
\succsim''_{n+1}&:\{\{z\}\}\succ\Sigma^{n+1}_1\succ\dots\succ\Sigma^{n+1}_{l_{n+1}}.
\end{align*}
By CUT, $R_{\pi_1}=R_{\pi_2}$. So, $xI_{\pi_2}y$.

Let $\pi_3=\mset{\succsim'''_1,\dots,\succsim'''_{n+1}}$, where $\succsim_{i}''':\{\{z\}\}$ for $1\leq i\leq n+1$. By NT, we have $xI_{\pi_3}y$. 

Notice that $\pi_2\in \pi_3\oplus\pi$. Hence, CON implies that $R_{\pi_2}\mid_{\{x,y\}}=R_{\pi}\mid_{\{x,y\}}$. Since we have already confirmed that $xI_{\pi_2}y$, we can conclude that $xI_{\pi}y$. This completes the induction step. 
\end{proof}

\begin{proof}[Proof of Theorem \ref{theorem:CUT and UCON}]

The `if' part is straightforward. We will prove the `only if' part. Let $R$ be an SRS satisfying CUT, UCON, and NT. 

The proof proceeds in three steps.

\textbf{Step 1.} We will prove Claim \ref{claim:single-class unanimous} in the proof of Theorem \ref{theorem:CUT and CON}. 

Let $\succsim\in\mathcal{R}$ be any single-class power relation such that $\succsim:\Sigma$. Fix any distinct $x,y\in X$. Suppose that there exists $S\subseteq X\setminus\{x,y\}$ such that $S\cup\{x\}\in\Sigma$ and $S\cup\{y\}\notin \Sigma$. 

Let $\pi_{1}:=\mset{\succsim}$. 

Let $\pi_2:=\mset{\succsim_1}$, where $\succsim_1:\{S\cup\{x\},S\cup\{y\}\}$. NT says that $xI_{\pi_2}y$. 

Let $\pi_3:=\mset{\succsim,\succsim_1}$. By UCON, we have $R_{\pi_1}\mid_{\{x,y\}}=R_{\pi_3}\mid_{\{x,y\}}$. 

Let $\pi_4:=\mset{\succsim_2}$, where $\succsim_2:\Sigma\cup \{S\cup\{y\}\}$. CUT implies $R_{\pi_3}=R_{\pi_4}$. To conclude, we have that $R_{\pi_1}\mid_{\{x,y\}}=R_{\pi_4}\mid_{\{x,y\}}$. 

According to the above argument, by adding some elements to $\Sigma$, we can find $\succsim':\Sigma'$, where $\Sigma'\supseteq \Sigma$, such that $R_{\pi}\mid_{\{x,y\}}=R_{\pi'}\mid_{\{x,y\}}$ and $S\cup\{x\}\in \Sigma'\Leftrightarrow S\cup \{y\}\in \Sigma'$ for all $S\subseteq X\setminus \{x,y\}$ (i.e., $\succsim'$ is symmetric with respect to $x$ and $y$). Hence, NT says that $xI_{\pi'}y$, and so we obtain that $xI_{\pi}y$. This proves Claim \ref{claim:single-class unanimous}. 

\vspace{0.5em}

\textbf{Step 2.} We will prove that for any $\succsim\in\mathcal{R}$ that has exactly two equivalence classes, i.e., $\succsim:\Sigma_1\succ\Sigma_2$, we have that $R_{\succsim}=R^{U}_{\succsim}$. 

Fix any $z,w\in X$ that do not appear in $\pi$. 

Let $\pi_1:=\mset{\succsim_1}$, where $\succsim_1:\Sigma_1\cup\{\{z\}\}$. By Step 1, $xI_{\pi_1}y$.

Let $\pi_2:=\mset{\succsim_2}$, where $\succsim_2:\{\{z\}\}\succ\{\{w\}\}$. By NT, $xI_{\pi_2}y$.

Let $\pi_3:=\mset{\succsim_3}$, where $\{\{w\}\}\sim\Sigma_2$. By Step 1, $xI_{\pi_3}y$. 

Let $\pi_4:=\mset{\succsim_1,\succsim_2,\succsim_3}$. By UCON, we have $xI_{\pi_4}y$. 

Let $\pi_5:=\mset{\succsim_4}$, where $\succsim_4:\Sigma_1\cup\{\{z\}\}\succ\{\{w\}\}\sim\Sigma_2$. By CUT and $xI_{\pi_4}y$, we have $xI_{\pi_5}y$.

Let $\pi_6:=\mset{\succsim_1,\succsim_5,\succsim}$, where $\succsim_5:\Sigma_2\cup\{\{w\}\}$. By CUT again, we have $xI_{\pi_6}y$.

By Step 1, we have $xI_{\mset{\succsim_1}}y$ and $xI_{\mset{\succsim_5}}y$. Therefore, UCON says that $R_{\pi_6}\mid_{\{x,y\}}=R_{\mset{\succsim}}$. To conclude, we have that $xI_{\mset{\succsim}}y$. 

\vspace{0.5em}

\textbf{Step 3.} We will prove the theorem. Take any $\pi\in \Pi$. Using CUT repeatedly, we obtain $\pi'=\mset{\succsim_1,\dots,\succsim_n}$ such that $R_{\pi'}=R_{\pi}$ and each power relation $\succsim_i$ in $\pi'$ has at most two equivalence classes. Steps 1 and 2 imply that $xI_{\succsim_i}y$ for all $i\in[n]$. Thus, UCON says that $xI_{\pi'}y$. Equivalently, we obtain $xI_{\pi}y$.
\end{proof}

\begin{proof}[Proof of Theorem \ref{theorem:CON and UCON}]
Suppose to the contrary that there exists an SRS $R$ that satisfies CON, UCON, and LE. 

The proof proceeds in two steps. 

\textbf{Step 1.} We will prove the following claim. 

\begin{claim}
\label{claim:symmetric one-class}
Let $x,y\in X$, $S\in\mathcal{X}$ with $S\subseteq X\setminus\{x,y\}$, $\succsim_1:\{S\cup\{x\}\}$ and $\succsim_{2}:\{S\cup\{y\}\}$. Then, $xI_{\mset{\succsim_1,\succsim_2}}y$.  
\end{claim}

Suppose to the contrary that the claim does not hold. Then, there exist $x,y\in X$, $S\in\mathcal{X}$ with $S\subseteq X\setminus\{x,y\}$, $\succsim_1:\{S\cup\{x\}\}$, and $\succsim_{2}:\{S\cup\{y\}\}$ such that $xP_{\pi_1}y$, where $\pi_1:=\mset{\succsim_1,\succsim_2}$. 

Let $\pi_2:=\mset{\succsim'_1,\succsim'_2}$, where $\succsim'_1:\{S\cup\{x\},S\cup\{y\}\}$ and $\succsim'_2:=\succsim_1$. Since $\pi_2\in\pi_1\oplus\pi_1$, CON implies that $xP_{\pi_2}y$. However, $xI_{\succsim'_1}y$ and $xI_{\succsim'_2}y$ should hold by LE. Therefore, UCON implies that $xI_{\pi_2}y$. This contradicts $xI_{\succsim'_2}y$. This completes the proof of Claim \ref{claim:symmetric one-class}.

\vspace{0.5em}

\textbf{Step 2.} Let $\pi_1:=\mset{\succsim_1,\succsim_2}$, $\pi_2:=\{\succsim'_1,\succsim'_2\}$, and $\pi:=\mset{\sqsupseteq_1,\sqsupseteq_2}$, where
\begin{align*}
\pi_1&:\begin{cases}
\succsim_1:\{\{x\}\}\\
\succsim_2:\{\{y\}\}
\end{cases}\\
\pi_2&:\begin{cases}
\succsim'_1:\{\{y,z\}\}\\
\succsim'_2:\{\{x,z\}\}
\end{cases}\\
\pi&:\begin{cases}
\sqsupseteq_1:\{\{x\}\}\succ\{\{y,z\}\}\\
\sqsupseteq_2:\{\{y\},\{x,z\}\}.
\end{cases}
\end{align*}
Using Step 1, we have that $xI_{\pi_1}y$ and $xI_{\pi_2}y$. Hence, CON implies that $xI_{\pi}y$ (because $\pi\in\pi_1\oplus\pi_2$). However, as $xP_{\mset{\sqsupseteq_1}}y$ and $xI_{\mset{\sqsupseteq_2}}y$ by LE, it follows that $xP_{\pi}y$ by UCON. This contradicts $xI_{\pi}y$.  
\end{proof}

\begin{proof}[Proof of Theorem \ref{theorem:LE and WCUT}]

Suppose to the contrary that there exists an SRS $R$ that satisfies LE, WCUT, and INDS.

Let 
\[\pi_1:\begin{cases}
\succsim_1:\{\{x\}\}\succ\{\{z\}\}\succ\{\{y\}\}.
\end{cases}\]
Then, $xP_{\pi_1}y$ by LE.
\[\pi_2:\begin{cases}
\succsim_2:\{\{x\}\}\succ\{\{z\}\}\\
\succsim_3:\{\{z\}\}\succ\{\{y\}\}.
\end{cases}\]
WCUT says that $R_{\pi_1}=R_{\pi_2}$. Hence, $xP_{\pi_2}y$.

Let 
\[\pi_3:\begin{cases}
\succsim_4:\{\{z\}\}\succ\{\{x\}\}\\
\succsim_5:\{\{y\}\}\succ\{\{z\}\}.
\end{cases}
\]
Using INDS, we have $xP_{\pi_3}y$.

Let 
\[
\pi_4:\begin{cases}
\succsim_6:\{\{y\}\}\succ\{\{z\}\}\succ\{\{x\}\}.
\end{cases}
\]
WCUT says that $R_{\pi_3}=R_{\pi_4}$. Hence, $xP_{\pi_4}y$. But LE demands that $yP_{\pi_4}x$. Contradiction.

\end{proof}

\begin{proof}[Proof of Theorem \ref{theorem:LE and CON}]

Suppose to the contrary that there exists an SRS $R$ that satisfies LE, CON, and ISPR. 

Let $x,y\in X$ be distinct individuals. Let $\succsim_1:\{\{x\}\}$ and $\succsim_{2}:\{\{y\}\}$, and $\pi:=\mset{\succsim_1,\succsim_2}$. Suppose that $xR_{\pi}y$. 

Now, let $\pi':=\mset{\succsim_1',\succsim_2'}$, where $\succsim'_1:\{\{y\}\}\succ\{\{x\}\}$ and $\succsim_2':\{\{x\},\{y\}\}$. Since $\pi'\in\pi\oplus\pi$, we have $xR_{\pi'}y$ by CON. 

By ISPR, we also obtain that $xR_{\mset{\succsim'_1}}y$. But LE says that $yP_{\mset{\succsim'_1}}x$. This contradiction follows from our assumption that $xR_{\succsim}y$. However, even if we assume that $yR_{\succsim}x$, a parallel argument also yields a contradiction. As $R_{\succsim}$ must be complete by definition, this means that there should be no SRS satisfying the three axioms.
\end{proof}

\begin{proof}[Proof of Theorem \ref{theorem:LE and UCON}]
Suppose to the contrary that there exists an SRS $R$ that satisfies LE, UCON, INDS, and TA. 

Fix any $x_1,x_2,x_3,x_4,x_5\in X$ and let $X_0:=\{x_1,x_2,x_3,x_4,x_5\}$. For each pair of distinct $x,y\in X_0$, fix any $\succsim^x_y\in \mathcal{R}(\{x_1\},\{x_2\},\{x_3\},\{x_4\},\{x_5\})$ such that $\{x\}\succ^x_y\{y\}$. 

We define a binary relation $\sqsupseteq$ as follows. 
\begin{equation}
\label{equation:supseteq}
x\sqsupseteq y\Leftrightarrow xR_{\mset{\succsim^x_y,\succsim^y_x}}y.
\end{equation}
Notice that $\succsim^x_y$ is not unique, but if two power relations $\succsim^x_y,\bar{\succsim}^x_y$ both belong to $\mathcal{R}(\{x_1\},\{x_2\},\{x_3\},\{x_4\},\{x_5\})$ and $\{x\}\succ^x_y\{y\}$ and $\{x\}\bar{\succ}^x_y\{y\}$, then $\succsim^x_y$ and $\bar{\succsim}^x_y$ can be transformed into each other by shifting the positions of $\{x,y\}$-nondiscriminating sets. Hence, INDS says that equation \eqref{equation:supseteq} is well-defined. We denote its asymmetric part by $\sqsupset$ (i.e., $x\sqsupset y\Leftrightarrow (x\sqsupseteq y\text{ and }\neg(y\sqsupseteq x))$) and its symmetric part by $\simeq$ (i.e., $x\simeq y\Leftrightarrow (x\sqsupseteq y\text{ and }y\sqsupseteq x)$). 

\vspace{0.5em}

\textbf{Case 1}: There exist $x,y,z\in X_0$ such that $x\simeq y$, $y\simeq z$, and $z\simeq x$. 

In this case, let $p,q$ be the elements of $X_0$ other than $x$, $y$, and $z$, and let
\[\pi_1:\begin{cases}
\succsim_1:\{\{x\}\}\succ\{\{y\}\}\succ\{\{z\}\}\succ\{\{p\}\}\succ\{\{q\}\},\\
\succsim_2:\{\{y\}\}\succ\{\{z\}\}\succ\{\{x\}\}\succ\{\{p\}\}\succ\{\{q\}\},\\
\succsim_3:\{\{z\}\}\succ\{\{x\}\}\succ\{\{y\}\}\succ\{\{p\}\}\succ\{\{q\}\}.
\end{cases}
\]

Then, we have that $xI_{\mset{\succsim_1,\succsim_2}}y$ by $x\simeq y$ and INDS. We also have that $xP_{\mset{\succsim_3}}y$ by LE. Hence, UCON says that $xP_{\pi_1}y$. 

Similarly, we can also derive that $yP_{\pi_1}z$ and $zP_{\pi_1}x$. These contradict TA.

\vspace{0.5em}

\textbf{Case 2}: There exist $x,y,z\in X_0$ such that $x\sqsupset y$ and $y\sqsupset z$. 

Let $\pi_2:=\mset{\succsim_2,\succsim_3}$, where $\succsim_2$ and $\succsim_3$ are the power relations defined in Case 1. 
Then, $xP_{\pi_2}y$ by $x\sqsupset y$ and INDS; $yP_{\pi_2}z$ by $y\sqsupset z$ and INDS; $zP_{\pi_2}x$ by LE and UCON. These contradict TA. 

\vspace{0.5em}

\textbf{Case 3}: Suppose that neither Case 1 nor Case 2 holds. That is, 
\begin{align}
\label{equation:case1}
&\text{There is no }x,y,z\in X_0\text{ such that }x\simeq y,y\simeq z,\text{ and }z\simeq x;\\
\label{equation:case2}
&\text{There is no }x,y,z\in X_0
\text{ such that }x\sqsupset y\text{ and }y\sqsupset z.\end{align}
By \eqref{equation:case1}, there exist $x,y\in X_0$ such that $x\sqsupset y$. Without loss of generality, we can assume that $x=x_1$ and $y=x_2$. 

Let 
\begin{align*}
X^1_0:=\{w\in X_0\mid x_1\sqsupset w\},\\
X^2_0:=\{w\in X_0\mid x_1\simeq w\},\\
X^3_0:=\{w\in X_0\mid w\sqsupset x_1\}.
\end{align*}

By \eqref{equation:case2}, we can conclude that $X^3_0=\emptyset$ (otherwise, say $x_3\in X^3_0$, then it follows that $x_3\sqsupset x_1$ and $x_1\sqsupset x_2$, which contradict \eqref{equation:case2}). 

By definition, we have $X_0=X^1_0\cup X^2_0\cup X^3_0$ and $\lvert X\vert =5$. So, given that $X^3_0=\emptyset$, it follows that
\begin{equation}
\label{equation:at least three}
\min\{\lvert X^1_0\rvert,\lvert X^2_0\rvert\}\geq 3.
\end{equation}
\begin{itemize}
\item Suppose $\lvert X^1_0\rvert \geq 3$. We can verify that $u\simeq v$ for all $u,v\in X^1_0$ as follows. If there exist some $u,v\in X^1_0$ such that $u\sqsupset v$, then it follows that $x_1\sqsupset u$ ($\because u\in X^1_0$) and $u\sqsupset v$, which contradict \eqref{equation:case2}. Therefore, with $\lvert X^1_0\rvert\geq 3$ in mind, we can conclude that there exist $u,v,w\in X^1_0$ such that $u\simeq v$, $v\simeq w$, and $w\simeq u$. But this also contradicts \eqref{equation:case1}.
\item Suppose $\lvert X^2_0\rvert \geq 3$. By \eqref{equation:case2}, there must exist $u,v\in X^2_0$ such that $u\sqsupset v$ (otherwise, there must exist $u,v,w\in X^2_0$ such that $u\sqsupset v,v\sqsupset w$, and $w\sqsupset u$, contradicting \eqref{equation:case2}). As $u\in X^2_0$ (this means $x_1\simeq u$), $x_1$ and $v$ must be distinct. 
\begin{itemize}
\item If $v\sqsupset x_2$, then it follows that $u\sqsupset v$ and $v\sqsupset x_2$, which contradicts \eqref{equation:case2}.
\item If $v\simeq x_2$, then we have $v\simeq x_1$, $x_1\sqsupset x_2$, and $x_2\simeq v$. Let 
\[\pi_3:\begin{cases}
\succsim_6&:\{\{x_1\}\}\succ\{\{x_2\}\}\succ\{\{v\}\},\\
\succsim_7&:\{\{x_2\}\}\succ\{\{v\}\}\succ\{\{x_1\}\},\\
\succsim_8&:\{\{v\}\}\succ\{\{x_1\}\}\succ\{\{x_2\}\}.
\end{cases}
\]
Then, $x_1 I_{\mset{\succsim_6,\succsim_7}}v$ by $x_1\simeq v$ and INDS, and $vP_{\mset{\succsim_8}}x_1$ by LE. So, UCON implies that $vP_{\pi_3}x_1$. We also have that $x_1 P_{\mset{\succsim_6,\succsim_7}}x_2$ by $x_1\sqsupset x_2$ and INDS, and $x_1 P_{\succsim_8} x_2$ by LE. So, UCON implies that $x_1 P_{\pi_3}x_2$. Furthermore, we have that $x_2 I_{\mset{\succsim_7,\succsim_8}}v$ by $x_2\simeq v$ and INDS, and that $x_2 P_{\mset{\succsim_6}}v$ by LE. Hence, UCON says that $x_2 P_{\pi_3} v$. To sum up, we obtain $vP_{\pi_3}x_1$, $x_1 P_{\pi_3}x_2$, and $x_2 P_{\pi_3}v$. This contradicts TA.
\end{itemize}
\end{itemize}

Now, we have derived a contradiction in all possible cases (Case 1, Case 2, and Case 3). This completes the proof of the theorem.
\end{proof}

\subsection{Proof of Theorem \ref{theorem:characterization of PEL}}

We say that an SRS $R$ satisfies \textit{independence of Lex-cel-indifferent relations} (ILIR) if for any $\pi\in \Pi$, $x,y\in X$, and $\succsim\in \mathcal{R}$ such that $xI^{L}_{\succsim}y$, we have that $R_{\pi}\mid_{\{x,y\}}=R_{\pi+\mset{\succsim}}\mid_{\{x,y\}}$.

\begin{lemma}
\label{lemma:ILIR}
If an SRS $R$ satisfies CA and ISPR, then $R$ also satisfies ILIR.
\end{lemma}

\begin{proof}
Let $\succsim\in \mathcal{R}$ with $\succsim:\Sigma_1\succ\Sigma_2\succ\dots\succ\Sigma_l$. Let $x,y\in X$. Assume that $xI^{L}_{\succsim}y$.

For each $k\in[l]$, we have 
\[t_k:=\lvert \{S\in \Sigma_k\mid x\in S,y\notin S\}\rvert=\lvert \{S\in \Sigma_k\mid y\in S,x\notin S\}.\]
Hence, using $A_{k,i},B_{k,i}\subseteq X\setminus\{x,y\}$ ($k\in[l]$ and $i\in[t_k]$), we can label the elements as follows: for each $k\in[l]$, let
\begin{align*}
\{S\in\Sigma_k\mid x\in S,y\notin S\}=\{A_{k,1}\cup\{x\},A_{k,2}\cup\{x\},\dots,A_{k,t_k}\cup\{x\}\},\\
\{S\in\Sigma_k\mid y\in S,x\notin S\}=\{B_{k,1}\cup\{y\},B_{k,2}\cup\{y\},\dots,B_{k,t_k}\cup\{y\}\}.
\end{align*}

Take $z_{i,j}\in X\setminus \bigcup (\Sigma_1\cup\dots\cup\Sigma_l)$ for $1\leq i\leq l$ and $1\leq j\leq t_i$. 

Let $\rho:\mathcal{X}\rightarrow\mathcal{X}$ be such that 
\begin{align*}
A_{i,j}\cup\{x\}&\mapsto\{z_{i,j},x\}\text{ for }1\leq i\leq l,1\leq j\leq t_i\\
B_{i,j}\cup\{y\}&\mapsto\{z_{i,j},y\}\text{ for }1\leq i\leq l,1\leq j\leq t_i\\
C&\mapsto C\text{ for any other }C
\end{align*}
Let $\succsim':=\succsim^\rho$. By the construction of $\succsim$, the $k$-th equivalence class of $\succsim'$ contains $\{z_{k,1},x\},\dots,\{z_{k,t_k},x\},\{z_{k,1},y\},\dots,\{z_{k,t_k},y\}$ and the other coalitions contain either (i) both $x$ and $y$ or (ii) neither $x$ nor $y$. Thus, $\succsim'$ is symmetric with respect to $x$ and $y$.

Now, take any $\pi\in \Pi$. We have that 

\begin{align*}
&R_{\pi+\mset{\succsim}}\mid_{\{x,y\}}\\
=&R_{(\pi+\succsim)^\rho}\mid_{\{x,y\}}\quad\because \text{ CA}\\
=&R_{\pi^\rho+\succsim^\rho}\mid_{\{x,y\}}\\
=&R_{\pi^\rho}\mid_{\{x,y\}}\quad\because\text{ ISPR}\\
=&R_{\pi}\mid_{\{x,y\}}\quad\because\text{ CA}.
\end{align*}
\end{proof}

For $\pi\in \Pi$ with $\pi=\mset{\succsim_1,\dots,\succsim_n}$ and $x,y\in X$, let $P(x,y;\pi):=\sum_{\substack{i\in[n]:\\xP^{L}_{\succsim_i}y}}\mset{\succsim_i}$.

\begin{lemma}
\label{lemma:contraction}
Let $R$ be an SRS satisfying CA and QCON.
Let $\pi\in \Pi$ and $x,y\in X$. If $\vert P(x,y;\pi)\rvert,\lvert P(y,x;\pi)\rvert>0$ and $xP_{\pi}y$, then there exists $\pi'\in \Pi$ such that $\lvert P(x,y;\pi')\rvert<\lvert P(x,y;\pi)\rvert$, $\lvert P(y,x;\pi)\rvert>0$, and $xP_{\pi'}y$.
\end{lemma}

\begin{proof}[Proof of Theorem \ref{theorem:characterization of PEL}]
Let $\pi=\mset{\succsim_1,\dots,\succsim_n}$ and $x,y\in X$. Assume that $a:=\lvert P(x,y;\pi)\rvert>0$, $b:=\lvert P(y,x;\pi)\rvert >0$, and $xP_{\pi}y$.

Without loss of generality, we can assume that $P(x,y;\pi)=\mset{\succsim_1,\dots,\succsim_a}$ and $P(y,x;\pi)=\mset{\succsim_{a+1},\dots,\succsim_{a+b}}$. 

Choose $z\in X$ that does not appear in $\pi$. Construct $\pi'=\mset{\succsim_1',\dots,\succsim_n'}\in \Pi$ in such a way that for each $i\in[n]$, $\succsim_i'$ is obtained from $\succsim_i$ by adding $z$ to each of the coalitions ranked in $\succsim_i$ (e.g., if $\succsim_i:\{\{x\}\}\succ\{\{x,y\},\{y\}\}$, then $\succsim_i':\{\{x,z\}\}\succ\{\{x,y,z\},\{y,z\}\}$). By CA, $R_{\pi'}\mid_{\{x,y\}}=R_{\pi}\mid_{\{x,y\}}$. Hence, $xP_{\pi'}y$ also holds. Let $\succsim_i:\Sigma^i_1\succ\dots\succ\Sigma^i_{l_i}$ and $\succsim'_i:\Gamma^i_i\succ\dots\succ\Gamma^i_{l_i}$ for each $i\in[n]$.

We define $\pi''=\mset{\succsim_1'',\dots,\succsim_n''}\in \pi\oplus\pi'$ by taking 

\[\rho(i)=\begin{cases}
a+i\quad&\text{if }1\leq i\leq b,\\
i-b&\text{if }b+1\leq i\leq a+b,\\
i&\text{if }a+b+1\leq i\leq n,
\end{cases}\]
and letting
\begin{align*}
\succsim''_i&:\Gamma^{a+i}_{1}\succ\dots\succ\Gamma^{a+i}_{l_{a+i}}\succ\Sigma^i_1\succ\dots\succ\Sigma^i_{l_i}\text{ if }1\leq i\leq b,\\
\succsim''_i&:\Sigma^i_1\succ\dots\succ\Sigma^i_{l_i}\succ\Gamma^{i-b}_1\succ\dots\succ\Gamma^{i-b}_{l_{i-b}} \text{ if } b+1\leq i\leq n.
\end{align*}
Then, we can verify that $yP_{\succsim_i''}x$ for every $i$ such that $1\leq i\leq b$ and $i=a+b$, and $xI_{\succsim''_i}y$ for every $i$ such that $a+b+1\leq i\leq n$. By QCON, we have $xP_{\pi''}y$. 

Hence, $P(x,y;\pi'')\subseteq \mset{\succsim''_{b+1},\dots,\succsim_{a+b-1}''}$. This means that $\lvert P(x,y;\pi'')\rvert\leq a-1<\lvert P(x,y;\pi)\rvert$. Furthermore, $\lvert P(y,x;\pi'')\rvert>0$ also holds. Therefore, this profile $\pi''$ is one instance of the required profile.
\end{proof}

\begin{proof}[Proof of Theorem \ref{theorem:characterization of PEL}]
The `if' part is straightforward. We will prove the `only if' part. Let $R$ be an SRS satisfying the five axioms. By Lemma \ref{lemma:ILIR}, $R$ satisfies ILIR as well. Let $\pi=\mset{\succsim_1,\dots,\succsim_n}$. 

\textbf{Case 1}: Suppose that $xR^{L}_{\succsim_i}y$ for all $i\in[n]$ and $xP^{L}_{\succsim_j}y$ for at least one $j\in[n]$. 

Let $\pi'$ be obtained from $\pi$ by dropping all $\succsim_i$ such that $xI^{L}_{\succsim_i}y$. By ILIR, we have $R_{\pi'}\mid_{\{x,y\}}=R_{\pi}\mid_{\{x,y\}}$. As $xP^{L}_{\succsim}y$ for all $\succsim\in \pi'$, LE and QCON imply that $xP_{\pi'}y$. Therefore, we can conclude that $xP_{\pi}y$.

\vspace{0.5em}

\textbf{Case 2}: Suppose that $xI^{L}_{\succsim_i}y$ for all $i\in[n]$.

Let $\pi':=\mset{\succsim_1}$. By ILIR, we have $R_{\pi'}\mid_{\{x,y\}}=R_{\pi}\mid_{\{x,y\}}$. As $xI_{\pi'}y$ by LE, we have $xI_{\pi}y$ as well. 

\vspace{0.5em}

\textbf{Case 3}: $yR^{L}_{\succsim_i}x$ for all $i\in[n]$ and $yP^{L}_{\succsim_j}x$ for at least one $j\in[n]$.

Just as in Case 1, we can verify that $yP_{\pi}x$ in \textbf{Case 3}.

\vspace{0.5em}

\textbf{Case 4}: There exist $i,j\in[n]$ such that $xP^{L}_{\succsim_i}y$ and $yP^{L}_{\succsim_j}x$.

We will prove that $xI_{\pi}y$ in this case. Suppose, for instance, that $xP_{\pi}y$. Applying Lemma \ref{lemma:contraction} repeatedly, we can reach $\pi'\in \Pi$ such that $\lvert P(x,y;\pi')\rvert=0$, $\lvert P(y,x;\pi')\rvert>0$, and $xP_{\pi'}y$. But this $\pi'$ must be an instance of Case 3, and so $yP_{\pi'}x$ must hold. This contradiction proves that $\neg (xP_{\pi}y)$. Swapping $x$ and $y$ in the above argument, we can also confirm that $\neg (yP_{\pi}x)$. So, $xI_{\pi}y$ holds.
\end{proof}

\section{Independence of the axioms}

\subsection{Independence of the axioms in Section \ref{section:multiple inter-profile axioms}}
\label{appendix:CUT+CON/UCON}

Throughout this section, we fix a linear order\footnote{A binary relation $\succsim\in A\times A$ is called antisymmetric if $[a\succsim b\text{ and }b\succsim a]\Rightarrow a=b$. A \textit{linear order} is an antisymmetric weak order.} $\vartriangleright$ on $X$. We define an SRS $R^{\vartriangleright}$ as follows: for any $\pi\in \Pi$, $R_{\pi}=\vartriangleright$.

For $\pi\in \Pi$ and $x\in X$, let 
\[\sign_\pi(x):=\begin{cases}
1\quad&\text{if }x\text{ appears in }\pi\\
0\quad&\text{otherwise.}
\end{cases}\]
For $\pi\in \Pi$ and $x\in X$, let $\sign$
We define an SRS $R^S$ (\textit{sign rule}) as follows: for any $\pi\in \Pi$ and $x,y\in X$, we have $xR^S_\pi y\Leftrightarrow \sign_\pi(x)\geq \sign_{\pi}(y)$. In other words, the sign rule is very coarse: it only sees whether each individual appears (the sign of the number of its appearances). 

We define an SRS $R^{A}$ (\textit{appearance rule}) as follows: for any $\pi=\mset{\succsim_1,\dots,\succsim_n}\in \Pi$, where $\succsim_i\in \mathcal{R}(\mathcal{D}_i)$ for each $i\in[n]$, and $x,y\in X$, we have that 
\[xR^A_{\pi}y\Leftrightarrow a_{\pi}(x)\geq a_{\pi}(y),\]
where $a_{\pi}(z):=\sum_{i\in[n]}\lvert \{S\in \mathcal{D}_{i}\mid z\in S\}\rvert$ for each $z\in X$. In other words, $a_{\pi}(x)$ represents how many times $x$ appears in $\pi$. The appearance rule $R^A$ declares that $x$ is at least as good as $y$ if $x$ appears as many times as $y$. 

\begin{proposition}
\label{proposition:ind of CUT, CON, and NT}
The three axioms in Theorem \ref{theorem:CUT and CON} are logically independent. In fact, 
\begin{enumerate}
\item $R^{\vartriangleright}$ satisfies CUT and CON, but not NT.
\item $R^{S}$ satisfies CUT and NT, but not CON. 
\item $R^{A}$ satisfies CON and NT, but not CUT. 
\end{enumerate}
\end{proposition}

\begin{proof}
The first statement is straightforward. 

For the second statement, by the definition of a cut, it is easy to see that for any $\succsim\in \mathcal{R}$, its cut $(\succsim_1,\succsim_2)$, and any $x\in X$, we have that $x$ appears in $\mset{\succsim}$ if and only if $x$ appears in $\mset{\succsim_1,\succsim_2}$. Hence, CUT trivially holds under $R^S$. NT is also straightforward. But $R^S$ does not satisfy CON. Let 
\begin{align*}
\pi_1&:\begin{cases}
\succsim:\{\{x\}\}
\end{cases}\\
\pi_2&:\begin{cases}
\succsim':\{\{x,z\},\{y\}\}
\end{cases}\\
\pi&:\begin{cases}
\succsim'':\{\{x\},\{x,z\},\{y\}\}.
\end{cases}
\end{align*}

Then, we have that $\pi\in \pi_1\oplus\pi_2$, and that $xP^S_{\pi_1}y$, $xI^S_{\pi_2}y$, and $xI^{S}_{\pi}y$. This contradicts CON.

For the third statement, by the definition of a merging sum, it is easy to see that for any $\pi_1,\pi_2,\pi\in \Pi$ with $\pi\in \pi_1\oplus\pi_2$, and any $x\in X$, we have that $a_{\pi}(x)=a_{\pi_1}(x)+a_{\pi_2}(x)$. Hence, CON trivially holds. NT is also straightforward. But $R^A$ does not satisfy CUT. To see this, let 
\begin{align*}
\pi&:\begin{cases}
\succsim:\{\{x\}\}\succ\{\{y\}\}\succ\{\{z\}\}
\end{cases},\\
\pi'&:\begin{cases}
\succsim':\{\{x\}\}\succ\{\{y\}\},\\
\succsim'':\{\{y\}\}\succ\{\{z\}\}.
\end{cases}
\end{align*}

Then, $(\succsim',\succsim'')$ is a cut of $\succsim$. However, we have $xP^A_{\pi}y$ and $xI^A_{\pi'}y$, contradicting CUT (notice that $R^A$ does not even satisfy WCUT in this example). 
\end{proof}

\begin{proposition}
\label{proposition:ind CUT, UCON, NT}
The three axioms in Theorem \ref{theorem:CUT and UCON} are logically independent. In fact, 
\begin{enumerate}
\item $R^\vartriangleright$ satisfies CUT and UCON, but not NT.
\item $R^{S}$ satisfies CUT and NT, but not UCON. 
\item $R^{ML}$ satisfies UCON and NT, but not CUT.
\end{enumerate} \end{proposition}

\begin{proof}
The first statement is straightforward. 

For the second statement, we have also seen in Proposition \ref{proposition:ind of CUT, CON, and NT} that $R^S$ satisfies CUT and NT. But it does not satisfy UCON. To see this, let us focus on $\pi_1,\pi_2$ in the proof of Proposition \ref{proposition:ind of CUT, CON, and NT}. Then, we have $xP^S_{\pi_1}y$, $xI^S_{\pi_2}y$, but $xI^S_{\pi_1+\pi_2}y$, which contradict UCON.

For the third statement, $R^{ML}$ clearly satisfies UCON and NT. To see that $R^{ML}$ fails to satisfy CUT, let us focus on $\pi,\pi'$ in the proof of Proposition \ref{proposition:ind of CUT, CON, and NT}. Then, we have that $xP^{ML}_{\pi}y$ and $xI^{ML}_{\pi'}y$, which contradict CUT.
\end{proof}

We define an SRS $R^{L:U}$ as follows: for any $\pi\in \Pi$, 
\[R_\pi=\begin{cases}
L_{\pi}\quad&\text{if }\lvert \pi\rvert=1,\\
R^{U}_{\pi}\quad&\text{if }\lvert \pi\rvert\geq 2.
\end{cases}\]

\begin{proposition}
\label{proposition:ind LE, CON, UCON}
The three axioms in Theorem \ref{theorem:CON and UCON} are independent. In fact, 
\begin{enumerate}
\item $R^{L:U}$ satisfies LE and CON, but not UCON.
\item $R^{ML}$ satisfies LE and UCON, but not CON.
\item $R^{A}$ satisfies CON and UCON, but not LE.
\end{enumerate}
\end{proposition}

\begin{proof}
To confirm that each SRS satisfies the stated axioms is straightforward. We will only demonstrate why they fail to satisfy some axioms. 

For $R^{L:U}$, let $\pi:\begin{cases}\succsim:\{\{x\}\}\succ\{\{y\}\}\end{cases}$. Then, $xP^{L:U}_{\pi}y$ but $xI^{L:U}_{\pi+\pi}y$, which contradict UCON. 

For $R^{ML}$, let 
\begin{align*}
\pi_1&:\begin{cases}
\succsim_1:\{\{x\}\}\succ\{\{y\}\},\\
\succsim_2:\{\{x\}\}\succ\{\{y\}\},\\
\succsim_3:\{\{y\}\}\succ\{\{x\}\}.
\end{cases}\\
\pi_2&:\begin{cases}
\succsim_1':\{\{x,z\}\}\succ\{\{y,z\}\},\\
\succsim_2':\{\{y,z\}\}\succ\{\{x,z\}\},\\
\succsim_3':\{\{x,z\}\}\succ\{\{y,z\}\}.
\end{cases}\\
\pi_3&:\begin{cases}
\succsim_1'':\{\{x,z\}\}\succ\{\{y,z\}\}\succ\{\{x\}\}\succ\{\{y\}\}\\
\succsim''_2:\{\{y,z\}\}\succ\{\{x,z\}\}\succ\{\{x\}\}\succ\{\{y\}\}\\
\succsim''_3:\{\{y\}\}\succ\{\{x\}\}\succ\{\{x,z\}\}\succ\{\{y,z\}\}
\end{cases}
\end{align*}
Then, $\pi_3\in \pi_1\oplus\pi_2$ (for each $i\in[3]$, $\succsim''_i\in \succsim_i\oplus \succsim_i'$ holds). But we have $xP^{ML}_{\pi_1}y$, $xP^{ML}_{\pi_2}y$, and $yP^{ML}_{\pi_3}x$, which contradict QCON.

For $R^A$, with $\pi:\begin{cases}\succsim:\{\{x\}\}\succ\{\{y\}\}\end{cases}$, we have $xI^A_{\pi}y$. This contradicts LE.  
\end{proof}

\subsection{Independence of the axioms in Section \ref{section: further impossibilities}}

For $\succsim\in \mathcal{R}$, we define $\mathcal{S}(\succsim)\subseteq \mathcal{R}$ as the smallest set satisfying the following: 
\begin{itemize}
\item $\mset{\succsim}\in \mathcal{S}(\succsim)$
\item If $\pi=\mset{\succsim_1,\dots,\succsim_n}\in \mathcal{S}(\succsim)$ and $(\succsim_i',\succsim_i'')$ is a clear cut of $\succsim_i$, then $\pi':=\mset{\succsim_1,\dots,\succsim_{i-1},\succsim_i',\succsim_i'',\succsim_{i+1},\dots,\succsim_n}$ also belongs to $\mathcal{S}(\succsim)$. 
\end{itemize}

In other words, $\mathcal{S}(\succsim)$ is the set of all profiles that can be obtained by applying the ``clear cut'' to $\mset{\succsim}$ a finite number of times. Each element of $\mathcal{S}(\succsim)$ is called \textit{single-relational with origin $\succsim$}. We define $\mathcal{S}:=\bigcup_{\succsim\in\mathcal{R}}\mathcal{S}(\succsim)$. Each element of $\mathcal{S}$ is called \textit{single-relational}. 

For any 
\[
\pi:\begin{cases}
\succsim_1&:\Sigma^1_1\succ\dots\succ\Sigma^1_{l_1}\\
\succsim_2&:\Sigma^2_1\succ\dots\succ\Sigma^2_{l_2}\\
&\vdots\\
\succsim_n&:\Sigma^n_1\succ\dots\succ\Sigma^n_{l_n},
\end{cases}
\]
we define a directed graph called \textit{the neighboring graph of $\pi$}, denoted as $N(\pi)=(A(\pi),E(\pi))$, as follows:
\begin{itemize}
\item $A(\pi)$ is the set of nodes, defined as $A(\pi):=\{\Sigma^i_j\mid i\in[n],j\in[l_i]\}$, and
\item $E(\pi)$ is the set of edges, defined as $E(\pi):=\{(\Sigma,\Gamma)\mid \exists i\in[n]\text{ s.t. }\succsim_i:\dots\succ\Sigma\succ\Gamma\succ\dots\}$ (i.e., $E(\pi)$ is the set of all pairs $(\Sigma,\Gamma)$ such that $\Sigma$ and $\Gamma$ are placed adjacently in this order in some power relation $\succsim_i$ in $\pi$). 
\end{itemize}

We can observe two properties concerning the neighboring graph of $\pi$. 

\begin{lemma}
\label{lemma:neighboring graph}
We have the following:
\begin{enumerate}
\item If $\pi=\mset{\succsim_1,\dots,\succsim_n}\in \mathcal{S}(\succsim)$, $(\succsim_i',\succsim_i'')$ is a clear cut of $\succsim_i$, and $\pi'=\mset{\succsim_1,\dots,\succsim_{i-1},\succsim_i',\succsim_i'',\succsim_{i+1},\dots,\succsim_n}$, then $N(\pi)=N(\pi')$.
\item For any distinct $\succsim,\succsim'\in \mathcal{R}$, we have $N(\mset{\succsim})\neq N(\mset{\succsim'})$.
\end{enumerate}
\end{lemma}

\begin{proof}
Statement 1 is straightforward. So, we will prove only statement 2. Suppose $\succsim\neq \succsim'$, and let 
\begin{align*}
\pi&:\begin{cases}\succsim:\Sigma_1\succ\dots\succ\Sigma_l,\end{cases}\\
\pi'&:\begin{cases}\succsim':\Gamma_1\succ\dots\succ\Gamma_m.\end{cases}
\end{align*}
If $A(\pi)\neq A(\pi')$, then $N(\pi)\neq N(\pi')$ trivially holds. Suppose $A(\pi)=A(\pi')$ (i.e., $\{\Sigma_1,\dots,\Sigma_l\}=\{\Gamma_1,\dots,\Gamma_m\}$). In this case, we have 
\begin{align*}
E(\pi)&=\{(\Sigma_1,\Sigma_2),\dots,(\Sigma_{l-1},\Sigma_l)\},\\
E(\pi')&=\{(\Gamma_1,\Gamma_2),\dots,(\Gamma_{m-1},\Gamma_m)\}.
\end{align*}
If $\Sigma_1\neq \Gamma_1$, then there exists $\Delta\in A(\pi)$ such that $(\Delta,\Gamma_1)\in E(\pi)$, because $A(\pi)=A(\pi')$ demands that $\Gamma_1\in \{\Sigma_2,\dots,\Sigma_l\}$.  But we can confirm that $(\Delta,\Gamma_1)\notin E(\pi')$ for any $\Delta\in A(\pi')$ (because $\Gamma_1$ is the best class in $\succsim'$). Therefore, $E(\pi)\neq E(\pi')$.

Suppose $\Sigma_1=\Gamma_1$. Since $\succsim\neq \succsim'$, there must exist $\hat{k}$ ($2\leq \hat{k}\leq l$) such that $\Sigma_k=\Gamma_k$ for all $k<\hat{k}$ and $\Sigma_{\hat{k}}\neq\Gamma_{\hat{k}}$. Then, it follows that $(\Sigma_{\hat{k}-1},\Sigma_{\hat{k}})\in E(\pi)\setminus E(\pi')$. Therefore, $E(\pi)\neq E(\pi')$. 
\end{proof}

\begin{lemma}
\label{lemma:N is into}
For any $\pi\in \Pi$ and $\succsim,\succsim'\in \mathcal{R}$, if $\pi\in \mathcal{S}(\succsim)\cap \mathcal{S}(\succsim')$, then $\succsim=\succsim'$.
\end{lemma}

\begin{proof}
Assume $\pi\in \mathcal{S}(\succsim)\cap \mathcal{S}(\succsim')$. By statement 1 of Lemma \ref{lemma:neighboring graph}, we have $N(\pi)=N(\mset{\succsim})$ and $N(\pi')=N(\mset{\succsim'})$, respectively. So, $N(\mset{\succsim})=N(\mset{\succsim'})$. Using statement 2 of Lemma \ref{lemma:neighboring graph}, we can say that $\succsim=\succsim'$.
\end{proof}

Suppose that $\pi\in \mathcal{S}$ is a single-relational profile. Then, according to Lemma \ref{lemma:N is into}, there exists a unique $\succsim\in \mathcal{R}$ such that $\pi\in \mathcal{S}(\succsim)$. Hence, we call this $\succsim$ the \textit{origin} of $\pi$, denoted as $o(\pi)$. 

We define an SRS $R^{O}$ (\textit{origin rule}) as follows. For any $\pi\in \Pi$, let
\[R^{O}_{\pi}=\begin{cases}
L_{o(\pi)}\quad\text{if }\pi\in \mathcal{S},\\
R^U_{\pi}\quad\text{otherwise.}
\end{cases}\]

\begin{proposition}
The three axioms in Theorem \ref{theorem:LE and WCUT} are independent. In fact, 
\begin{enumerate}
\item $R^O$ satisfies LE and WCUT, but not INDS.
\item $R^{ML}$ satisfies LE and INDS, but not WCUT.
\item $R^{U}$ satisfies WCUT and INDS, but not LE.
\end{enumerate}
\end{proposition}

\begin{proof}
It is straightforward to verify that each SRS satisfies the first two axioms. We will only demonstrate why each fails to satisfy the last one.

For $R^O$, let us recall the profiles $\pi_2$ and $\pi_3$ in the proof of Theorem \ref{theorem:LE and WCUT}. We have that $xP^{O}_{\pi_2}y$, but $yP^{O}_{\pi_3}x$, which contradict INDS.

For $R^{ML}$, the counterexample is found in the proof of statement 3 of Proposition \ref{proposition:ind of CUT, CON, and NT} ($xP^{ML}_{\pi}y$ but $xI^{ML}_{\pi'}y$). 

$R^{U}$ trivially fails to satisfy LE.
\end{proof}

\begin{proposition}
The three axioms in Theorem \ref{theorem:LE and CON} are logically independent. In fact, 
\begin{enumerate}
\item $R^{L:U}$ satisfies LE and CON, but not ISPR.
\item $R^{ML}$ satisfies LE and ISPR, but not CON.
\item $R^{A}$ satisfies CON and ISPR, but not LE.
\end{enumerate}
\end{proposition}

\begin{proof}
It is straightforward to verify that each SRS satisfies the first two axioms. We will only demonstrate why each fails to satisfy the last one.

For $R^{L:U}$, let $\succsim:\{\{x\}\}\succ\{\{y\}\}$, $\succsim':\{\{x,y\}\}$, $\pi:=\mset{\succsim}$, and $\pi':=\mset{\succsim,\succsim'}$. Then, we have $xP^{L:U}_{\pi}y$ and $xI^{L:U}_{\pi'}y$. This contradicts ISPR. 

For $R^{ML}$ and $R^A$, the counterexamples are already shown in Proposition \ref{proposition:ind LE, CON, UCON}.
\end{proof}

\begin{proposition}
\label{proposition:ind LE,UCON,INDS,TA}
The four axioms in Theorem \ref{theorem:LE and UCON} are logically independent. In fact, 

\begin{enumerate}
\item $R^{ML}$ satisfies LE, UCON, and INDS, but not TA.
\item $R^{AL}$ satisfies LE, UCON, and TA, but not INDS.
\item $R^{L:U}$ satisfies LE, INDS, and TA, but not UCON.
\item $R^{A}$ satisfies UCON, INDS, and TA, but not LE.
\end{enumerate}
\end{proposition}

\begin{proof}
It is straightforward to confirm that each SRS satisfies the first three axioms. We will demonstrate why each fails to satisfy the last one. 

For $R^{ML}$, this has already been shown in Proposition \ref{proposition:lex-cel extensions}. 

For $R^{AL}$, let 
\begin{align*}
\pi:\begin{cases}
\succsim_1:\{\{z\}\}\succ\{\{x\}\}\succ\{\{y\}\},\\
\succsim_2:\{\{y\}\}\succ\{\{x\}\}.
\end{cases}\\
\pi':\begin{cases}
\succsim_1':\{\{x\}\}\succ\{\{y\}\}\succ\{\{z\}\}\\
\succsim_2
\end{cases}
\end{align*}
Then, we have $yP^{AL}_{\pi}x$ and $xI^{AL}_{\pi'}y$. This contradicts INDS.

The counterexamples for $R^{L:U}$ and $R^A$ are already shown in Proposition \ref{proposition:ind LE, CON, UCON}.
\end{proof}

\subsection{Independence of the axioms in Section \ref{section:characterization of PEL}}

For any $\succsim:\Sigma_1\succ\dots\succ\Sigma_l\in \mathcal{R}$ and $z\in X$, let 
\[\theta^s_{\succsim}(z):=\left(\sum_{\substack{S\in \Sigma_1:\\z\in S}}\frac{1}{\lvert S\rvert},\sum_{\substack{S\in \Sigma_2:\\z\in S}}\frac{1}{\lvert S\rvert},\dots,\sum_{\substack{S\in \Sigma_l:\\z\in S}}\frac{1}{\lvert S\rvert}\right).\]

We define $R^{PESL}$ (\textit{Pareto-extension split Lex-cel}) as follows. For any $\pi\in \Pi$ and $x,y\in X$, we have
\[
xR^{PESL}_{\pi}y
\Leftrightarrow\;
\neg\bigg(
\forall \succsim\in\pi,\,
\theta^s_{\succsim}(y)\geq^L
\theta^s_{\succsim}(x)
\quad\text{and}\quad
\exists \succsim\in\pi,\,
\theta^s_{\succsim}(y)>^L
\theta^s_{\succsim}(x)
\bigg).
\]

For any $\pi=\mset{\succsim_1,\dots,\succsim_n}\in \Pi$ with $\succsim_i\in \mathcal{R}(\mathcal{D}_i)$ for each $i\in[n]$ and $x,y\in X$, define $\epsilon_{\pi}(x,y)$ as follows: 
\[\epsilon_\pi(x,y):=\lvert \{i\in[n]\mid \lvert \mathcal{D}_i\rvert=1\text{ and }\lvert \left(\bigcup\mathcal{D}_i\right)\cap \{x,y\}\rvert=1\}\rvert.\]
We define $R^{PEL:U}$ as follows. For any $\pi\in \Pi$ and $x,y\in X$, 
\[R_{\pi}^{PEL:U}=\begin{cases}
R^{PEL}_\pi\quad\text{if }\epsilon_{\pi}(x,y)\leq 1\\
R^U_\pi\quad\text{if }\epsilon_{\pi}(x,y)\geq 2.
\end{cases}\]

\begin{proposition}
\label{proposition:ind ISPR,WCA,QUCON,QCON,LE}
The five axioms (LE, QCON, QUCON, CA, and ISPR) in Theorem \ref{theorem:characterization of PEL} are logically independent. In fact, 

\begin{enumerate}
\item $R^{WPL}$ satisfies all five axioms except ISPR.
\item $R^{PESL}$ satisfies all five axioms except WCA.
\item $R^{PEL:U}$ satisfies all five axioms except QUCON.
\item $R^{ML}$ satisfies all five axioms except QCON.
\item $R^A$ satisfies all five axioms except LE.
\end{enumerate}
\end{proposition}

\begin{proof}
It is straightforward to confirm that each SRS satisfies four of the five axioms. We will only demonstrate why each SRS fails to satisfy the remaining axiom.

For $R^{WPL}$, let $\succsim:\{\{x\}\}\succ\{\{y\}\}$, $\succsim':\{\{x,y\}\}$, $\pi:=\mset{\succsim}$, and $\pi':=\mset{\succsim,\succsim'}$. Then, we have $xP^{WPL}_{\pi}y$ and $xI^{WPL}_{\pi'}y$, which contradict ISPR. 

For $R^{PESL}$, let 
\begin{align*}
\pi&:\begin{cases}
\succsim_1:\{\{x\},\{y\}\}
\end{cases}\\
\pi'&:\begin{cases}
\succsim_2:\{\{x,z\},\{y\}\}.
\end{cases}
\end{align*}
Then, we have $xI^{PESL}_{\pi}y$ and $yP^{PESL}_{\pi'}x$, which contradict WCA.

For $R^{PEL:U}$, let $\pi:\begin{cases}\succsim:\{\{x\}\}\end{cases}$. Then, $xP_{\pi}^{PEL:U}y$ but $xI^{PEL:U}_{\pi+\pi}y$. This contradicts QUCON.

The counterexamples for $R^{ML}$ and $R^A$ are already shown in Proposition \ref{proposition:ind LE, CON, UCON}.
\end{proof}

\end{appendices}

\bibliography{library}

\end{document}